\documentclass[nofootinbib,preprint,amsmath,amssymb,showpacs,showkeys,aps]{revtex4-1}
\usepackage{graphicx}
\usepackage[utf8,latin1]{inputenc}
\usepackage{dcolumn}
\usepackage{mathrsfs}
\usepackage{subcaption}
\usepackage{gensymb}
\usepackage{tensor}
\usepackage{academicons}
\usepackage{mathtools, nccmath}
\usepackage[overload]{empheq}
\usepackage{tikz,xcolor}
\usepackage{cases}
\usepackage{setspace}
\usepackage{lipsum}
\usepackage[normalem]{ulem}
\usepackage{bm}
\usepackage{soul}
\usepackage{hhline}
\usepackage{bigints}
\usepackage{amsmath}
\usepackage[T1]{fontenc}
\usepackage{fancyhdr}
\newcommand{\qm}[1]{``#1''}
\usepackage{hyperref}
\hypersetup{colorlinks, linkcolor={red},citecolor={blue},urlcolor={blue}}  

\newcommand{\bsubeqs}{\begin{subequations}}
\newcommand{\esubeqs}{\end{subequations}}
\usepackage[all]{xy}
\usepackage{cancel}

\definecolor{lime}{HTML}{A6CE39}
\DeclareRobustCommand{\orcidicon}{
	\begin{tikzpicture}
	\draw[lime, fill=lime] (0,0) 
	circle [radius=0.16] 
	node[white] {{\fontfamily{qag}\selectfont \tiny ID}};
	\draw[white, fill=white] (-0.0625,0.095) 
	circle [radius=0.007];
	\end{tikzpicture}
	\hspace{-2mm}
}

\newcommand\R{\mathbb{R}}
\newcommand\C{\mathbb{C}}
\DeclareFontFamily{U}{mathx}{}
\DeclareFontShape{U}{mathx}{m}{n}{ <-> mathx10 }{}
\DeclareSymbolFont{mathx}{U}{mathx}{m}{n}
\DeclareFontSubstitution{U}{mathx}{m}{n}
\DeclareMathAccent{\widecheck}{0}{mathx}{"71}

\newcommand{\dd}{{\rm d}}
\newcommand{\ii}{{\rm i}}
\newcommand{\ee}{{\rm e}}

\fancypagestyle{plain}{%
  \fancyhf{}
  \fancyfoot[C]{\iffloatpage{}{\thepage}}
  }
\foreach \x in {A, ..., Z}{%
	\expandafter\xdef\csname orcid\x\endcsname{\noexpand\href{https://orcid.org/\csname orcidauthor\x\endcsname}{\noexpand\orcidicon}}
}

\makeatletter
\newcommand*\bigcdot{\mathpalette\bigcdot@{.7}}
\newcommand*\bigcdot@[2]{\mathbin{\vcenter{\hbox{\scalebox{#2}{$\m@th#1\bullet$}}}}}
\makeatother

\begin{document}

\title[Complex degenerate metrics in general relativity: \\ a covariant extension of the Moore-Penrose algorithm]{Complex degenerate metrics in general relativity: \\ a covariant extension of the Moore-Penrose algorithm}

\author{Arthur Garnier\orcidA{}$^{1}$} \email{arthur.garnier@math.cnrs.fr}
\author{Emmanuele Battista\orcidB{}$^{2}$}
\email{ebattista@lnf.infn.it}\email{emmanuelebattista@gmail.com}

\affiliation{$^1$ LAMFA, University of Amiens\\
$^2$ Istituto Nazionale di Fisica Nucleare, Laboratori Nazionali di Frascati, 00044 Frascati, Italy,}

\date{\today} 

\begin{abstract}

The Moore-Penrose algorithm provides a generalized notion of an inverse, applicable to degenerate  matrices. In this paper, we introduce a  covariant extension of the Moore-Penrose method that permits to deal with general relativity involving complex non-invertible metrics. Unlike the standard technique, this approach guarantees the uniqueness of the pseudoinverse metric through the fulfillment of a set of covariant relations, and it allows for the proper definition of a covariant derivative operator and  curvature-related tensors.  Remarkably, the degenerate nature of the metric can be given a geometrical representation in terms of a torsion tensor, which vanishes only in special cases. Applications of the new scheme to complex black hole geometries and cosmological models are also investigated, and a  generalized concept of geodesics that exploits the notion of autoparallel and extremal curves is presented. Relevance of our findings to quantum gravity and quantum cosmology is finally discussed.

\end{abstract}

\maketitle

\section{Introduction}

Complex metrics and complexified spacetimes are widely explored in gravity research. Historically, these concepts date back to the Lagrangian approach to quantum gravity, where the  otherwise oscillating path integral is managed by Wick rotating the time axis $90^{\degree}$ clockwise in the complex plane, thereby passing to the  Euclidean spacetime. The resulting integral is then exponentially damped although  still not potentially convergent, since the gravitational  Euclidean action can become arbitrarily negative  for metrics that are real on  the so-called Euclidean section, which is defined as the domain of the complexified spacetime where all  coordinates attain real values. This shortcoming,   referred to as conformal factor problem,   can be overcome by choosing the integration contour in the space of complex metrics, 
which gives better convergence properties (for details see Refs. \cite{Gibbons1978,Halliwell1989}).  Euclidean path integrals   have since become a valuable tool in various fields, such as black-hole thermodynamics \cite{GH1977},  gravitational instantons \cite{Hawking1976,Esposito1994} (see also Refs. \cite{Battista-Esposito2020,Battista-Esposito2022b} for an alternative geometric-measure-theory approach),  the Page curve and replica wormholes \cite{Almheiri2020,Penington2022}. 

The study of complex metrics also plays a pivotal role in describing topology-change processes \cite{Louko1995}, which are expected to be a key feature of quantum gravity.  This investigation has highlighted the need for criteria to select physically viable metrics, thereby ruling out the solutions leading to nonphysical results \cite{Louko1995,Kontsevich2021}.  Remarkably,  the so-called \qm{allowed} complex metrics can be singled out by applying the principles of quantum field theory in curved spaces, as   admissible models are those on which a consistent quantum field theory can be defined,  the underlying path integral being thus manifestly convergent  \cite{Witten2021,Lehners2021,Jonas2022}. The main conclusion is that acceptable geometries  satisfy a pointwise condition constraining the form of the complex metric components \cite{Louko1995,Kontsevich2021}, with the significant implication that the domain of sensible solutions is contractible to the space of Euclidean metrics.

One technique involving complex metrics worth mentioning is the Newman-Janis algorithm, which finds applications in general relativity and extended gravity theories \cite{Canonico2011}, especially in the realm of gravitational instantons \cite{Aksteiner2022,Lan2024} and scattering amplitudes \cite{Arkani-Hamed2019,Guevara2020}, and has recently been utilized in loop quantum gravity to construct a nonsingular rotating black hole spacetime \cite{Brahma2020}. This latter investigation furnishes a clear example that complex-valued metrics provide a pathway to avoid the singularities predicted by  Einstein theory. This feature is a key ingredient in the various  proposals for  quantum cosmology,  where complex metrics are often brought in.  Specifically, within the  Hartle-Hawking no-boundary paradigm,  it has been proved that the wave function of the universe in the semiclassical steepest-descent approximation can exhibit an oscillating behavior,   pointing to the occurrence of classical spacetimes, provided  the underlying integration contour is dominated by one or more saddle points where the metric is complex \cite{Halliwell1990}.  Interesting developments have emerged in this setting, following the recent introduction of a new procedure based on Lorentzian path integrals \cite{Feldbrugge2017a,Feldbrugge2017b,Dorronsoro2017,DiTucci2019}.  In this framework,  the contour of integration is deformed over metrics into the complex plane, and then  a complex-analysis-inspired method based on the Picard-Lefschetz theory of saddle point approximations is exploited to ensure the convergence of the  path integral \cite{Lehners2023}.  

Studies dedicated to averting singularities have also been pursued in classical gravity by utilizing the notion of complexified spacetimes.  These structures in fact permit to consistently implement in  classical cosmological settings the change-of-signature concept featuring the different approaches to the quantum cosmology problem \cite{Ellis1992a,Ellis1992b}. In these models, the  metric becomes degenerate at certain points and describes a singularity-free universe that smoothly transitions from a Euclidean to a Lorentzian regime at early times. Recently, this pattern has been extended also to static black-hole geometries \cite{Capozziello2024} (see also Ref. \cite{Ellis1997}), where the joint effect of a non-invertible metric at the event horizon and   an imaginary time variable leads to a  solution where the singularity at $r=0$ cannot be reached.

Degenerate (both complex-valued and real-valued) metrics appear in several gravity frameworks, ranging from classical contexts, such as bouncing cosmology (see e.g. Refs. \cite{Klinkhamer-Wang2019,Battista2020,Wang2021,Holdom2023a}), to more specialized settings, including the Ashtekar reformulation of general relativity \cite{Ashtekar1987} and first-order gravity \cite{Kaul2016,Kaul2016b,Kaul2018}, to quantum settings \cite{Tseytlin1981,Jacobson1987,Horowitz1990,Tucker1994,Jacobson1996}. Their presence introduces significant challenges in defining a consistent physical theory, and several strategies have been put forward to address these issues. In quantum gravity, classical geometries with extreme-curvature domains or non-invertible metrics are expected to be suppressed by  
enhanced quantum fluctuations arising from  the wavefunction of the universe \cite{Kiefer2007}. A similar idea is  also present   in loop quantum gravity, which envisages quantum corrections  smoothing out regions where the classical metric description would normally break down  \cite{Ashtekar2011,Ashtekar2021}. Another proposed technique relies on the use of regularization schemes  (see e.g. Ref. \cite{Louko1995}), which allow metrics becoming  degenerate at isolated points to be included in the  path integral, ensuring that they contribute meaningfully to the quantum gravitational dynamics \cite{Louko1995}.  At classical level, degeneracy problems are often resolved by imposing  junction conditions where the metric becomes non-invertible, which guarantee that Einstein equations  evolve smoothly through the spacetime \cite{Ellis1992a,Ellis1992b,Capozziello2024}. Continuous extensions have also proven useful, as discussed in Refs.  \cite{Klinkhamer2014,Klinkhamer2013a} (for details, see Ref.  \cite{Gunther}). In this framework, the field equations are defined  where the metric is non-invertible by means of  their limits evaluated over the spacetime points approaching the degeneracy region. A further valuable tool is represented by the Moore-Penrose algorithm, which permits to extend the theory via the so-called  pseudoinverse metric,  generalizing the usual notion of  inverse metric \cite{Cao2015,Gunther}. Despite its  effectiveness, the Moore-Penrose  method  suffers from a drawback, as the uniqueness of the pseudoinverse metric can only be obtained  by invoking  conditions that are not coordinate covariant and therefore not sensible in general relativity.

In this paper, we introduce a new perspective on degenerate metrics in general relativity by proposing a covariant extension of the Moore-Penrose scheme, which can be applied to complex metrics and is readily suitable for real-valued ones as well. Within this new  procedure, the uniqueness of the pseudoinverse metric is guaranteed by the application of a set of covariant relations. As a consequence, the ensuing curvature tensors  are  genuine tensors,  meaning that a well-defined extension of general relativity encompassing degenerate metrics can be formulated. The plan of the paper is thus as follows. In Sec. \ref{Sec:complexified-Schwarzschild-geom},   we  introduce the complexified Schwarzschild geometry, which represents the reference model for our investigation. The  complex-valued Schwarzschild metric is found to be degenerate,  and thus, in Sec. \ref{Sec:generalized-Moore-Penrose-algorithm}, we work out the generalized Moore-Penrose approach. Remarkably, we find that the metric degenerate character  has a geometrical counterpart in the presence of a torsion tensor, which vanishes only in highly symmetric models that will be discussed in the paper. In Sec. \ref{Sec:Geodesic-equations}, we explore a broader concept of geodesics that goes beyond the usual principles and is suitable for non-invertible geometries. Final remarks are reported in Sec. \ref{Sec:conclusions}. Supplementary material is provided in the appendices.

\emph{Conventions and notations.} We use  units $G=c=1$. Greek indices span four dimensions and range as $\alpha,\beta,\dots=0,\dots,3$, while Latin indices $a,b,\dots$ are five-dimensional and take values $0,\dots,4$. Round (respectively, square) brackets around tensor indices stands for the usual symmetrization (respectively, antisymmetrization) procedure, e.g., $A_{(ab)}=\frac{1}{2}(A_{ab}+A_{ba})$ (respectively, $A_{[ab]}=\frac{1}{2}(A_{ab}-A_{ba})$).

\section{The complexified Schwarzschild geometry} \label{Sec:complexified-Schwarzschild-geom}

In this section, we introduce the complexified Schwarzschild metric, which serves as our benchmark geometry, in the sense that it  allows us to better understand the concepts we will encounter in the paper. Our starting point will be the Euclidean Schwarzschild solution, which we discuss in  Sec. \ref{Sec:Euclidean-section} before  passing to the complex Schwarzschild geometry in Sec. \ref{Sec:Complex-Schwarzschild-1}.

\subsection{The Euclidean Schwarzschild geometry}\label{Sec:Euclidean-section}

Gravitational instantons are four-dimensional solutions of the Euclidean Einstein equations that are nonsingular on some section of the complexified spacetime \cite{Hawking1976,Esposito1994}. These geometries fulfill a crucial role in Euclidean quantum gravity, such as in the so-called one-loop approximation within the background-field method \cite{Esposito1994}. 

The Euclidean Schwarzschild metric is the most famous example of gravitational instanton. Expressed in  Schwarzschild coordinates $x^\mu=(\tau,r,\theta,\phi)$, it  reads  as 
\cite{GH1977,G1983,Esposito1994}
\begin{equation} \label{Schwarzschild-metric-1}
\dd s^2 = g_{\mu \nu} \dd x^\mu \dd x^\nu =   \left(1-\frac{2M}{ r}\right) \dd \tau^2  +  \left(1-\frac{2M}{ r }\right)^{-1} \dd r^2  + r^2 \left({\rm d}\theta^2+\sin^2\theta \, {\rm d} \phi^2\right),
\end{equation}
and the analytic continuation to the Lorentzian-signature metric 
\begin{align}
\dd s_{\rm L}^2  =  - \left(1-\frac{2M}{ r}\right) \dd t^2  +  \left(1-\frac{2M}{ r }\right)^{-1} \dd r^2  + r^2 \left({\rm d}\theta^2+\sin^2\theta \, {\rm d} \phi^2\right), 
\label{Lorentzian-1}
\end{align}
can be easily  obtained via the relation
\begin{align}
\tau = {\rm i}t. 
\label{tau-and-t}
\end{align}
The real Riemannian section of the complexified Schwarzschild spacetime  where $\tau$ is real is known as Euclidean section (the concept of sections of the complexified spacetime will be discussed in Sec. \ref{Sec:Complex-Schwarzschild-1} below). In this domain, it is possible to prove that  $\tau$ is a  periodic variable  with period equal to the inverse of the Hawking  temperature \cite{GH1977,G1983}.

If one adopts  Kruskal-Szekeres coordinates $X^\mu=(\xi,y,\theta,\phi)$, the metric (\ref{Schwarzschild-metric-1}) becomes
\begin{equation} \label{Schwarzschild-metric-2}
\dd s^2= \mathcal{G}_{\mu \nu} \dd X^\mu \dd X^\nu= \dfrac{32M^3}{r}\, {\rm e}^{-r/(2M)} \left({\rm d} \xi^2 
+{\rm d} y^2 \right)+ r^2 \left( {\rm d} \theta^2 
+ \sin^2 \theta \; {\rm d} \phi^2 \right),
\end{equation}
so that the corresponding Lorentzian metric 
\begin{align}
\dd s_{\rm L}^2=\dfrac{32M^3}{r}\, {\rm e}^{-r/(2M)} \left(-{\rm d} \zeta^2
+{\rm d} y^2  \right)+ r^2 \left( {\rm d} \theta^2  
+ \sin^2 \theta \; {\rm d} \phi^2  \right),    
\label{Lorentzian-2}
\end{align}
can be constructed if we write 
\begin{equation}
\xi = {\rm i} \zeta.
\label{xi-and-z}
\end{equation}
The  variables $\xi$ and $y$ are defined by 
\begin{subequations}
\label{y-and-xi-relations}
\begin{align}
& y+ {\rm i} \xi = {\rm e}^{{\rm i} \tau/(4M)} \sqrt{\dfrac{r}{2M}-1}  \; {\rm e}^{r/(4M)},
\\
& y = \cos \left(\dfrac{\tau}{4M}\right) \sqrt{\dfrac{r}{2M}-1} \; {\rm e}^{r/(4M)},
\end{align}    
\end{subequations}
and obey the relation
\begin{align}
\xi^2 + y^2 = \left(y + \ii \xi\right)\left(y - {\rm i} \xi\right)=\left(\dfrac{r}{2M}-1\right) \ee^{r/(2M)}.
\label{zeta-y-squared-relation}
\end{align}
In the Euclidean section,  $\xi$ and $y$ are real and hence   Eq. \eqref{zeta-y-squared-relation} constrains   the $r$ coordinate to satisfy
\begin{equation}\label{r-bigger-2M}
r \geq 2M.
\end{equation}
The consequences of the lower bound \eqref{r-bigger-2M} are twofold. First of all, it entails that the Euclidean Schwarzschild metric is positive-definite  in the  Euclidean section. Moreover, this domain  is geodesically complete, as the geodesics are not able to hit the singularity at $r=0$. As far as we know, this latter feature has never been demonstrated  explicitly in the literature. For this reason, after having derived the geodesic equations in Kruskal-Szekeres coordinates in Sec. \ref{Sec:motion-Euclidea-domain}, we provide an original proof of this result in Sec. \ref{Sec:geodesic-completeness}. This investigation prepares the ground for the study in Sec.   \ref{Sec:Geodesic-equations}, where we develop a generalized version of geodesic equations applicable to degenerate metrics.

\subsubsection{Motion equations} \label{Sec:motion-Euclidea-domain}

We now derive the geodesic equations for  the Euclidean Schwarzschild metric $\mathcal{G}_{\mu \nu}$ written in Kruskal-Szekeres coordinates $X^\mu$ (see Eq. \eqref{Schwarzschild-metric-2}). 

By spherical symmetry, we may assume that $\theta=\pi/2$ and $\dot{\theta}=0$, the dot denoting differentiation with respect to the affine parameter $\lambda$. The Lagrangian then reads as
\begin{align}
2\mathcal{L}=\mathcal{G}_{\mu \nu} \dot{X}^\mu \dot{X}^\nu=\frac{32 M^3\ee^{-r/(2M)}}{r}(\dot{\xi}^2+\dot{y}^2)+r^2\dot{\phi}^2.    
\label{Lagrangian-Euclidean-1}
\end{align}
Since $\phi$ is a cyclic variable, the Euler-Lagrange equation $\frac{\partial\mathcal{L}}{\partial\phi}=\frac{{\rm d}}{{\rm d}\lambda}\left(\frac{\partial\mathcal{L}}{\partial\dot{\phi}}\right)$ immediately yields a constant $J$ such that $\dot{\phi}=J/r^2$. Now, to obtain the equations satisfied by $\xi$ and $y$, these variables playing symmetric roles, we only need to treat the case of $\xi$, say. From Eq. \eqref{Lagrangian-Euclidean-1}, we  find 
\begin{align}
\frac{{\rm d}}{{\rm d}\lambda}\left(\frac{\partial\mathcal{L}}{\partial\dot{\xi}}\right)&=\frac{{\rm d}}{{\rm d}\lambda}\left(\frac{32M^3\dot{\xi}\ee^{-r/(2M)}}{r}\right)
\nonumber \\
&=\frac{32M^3\ddot{\xi}\ee^{-r/(2M)}}{r}+32M^3\dot{\xi}\left(\dot{\xi}\frac{\partial r}{\partial\xi}+\dot{y}\frac{\partial r}{\partial y}\right)\frac{\rm d}{{\rm d}r}\left(\frac{\ee^{-r/(2M)}}{r}\right),    
\end{align}
and 
\begin{align}
\frac{\partial\mathcal{L}}{\partial\xi}=16M^3(\dot{\xi}^2+\dot{y}^2)\frac{\partial r}{\partial\xi}\frac{\rm d}{{\rm d}r}\left(\frac{\ee^{-r/(2M)}}{r}\right)+8M^2\dot{\phi}^2\xi \ee^{-r/(2M)},    
\end{align}
and thus the Euler-Lagrange equation becomes
\begin{align}
\frac{32M^3\ddot{\xi}\ee^{-r/(2M)}}{r}&=\left[16M^3(\dot{\xi}^2+\dot{y}^2)\frac{\partial r}{\partial \xi}-32M^3\dot{\xi}\left(\dot{\xi}\frac{\partial r}{\partial \xi}+\dot{y}\frac{\partial r}{\partial y}\right)\right]\frac{\rm d}{{\rm d}r}\left(\frac{\ee^{-r/(2M)}}{r}\right)
\nonumber \\
&+8M^2\dot{\phi}^2\xi \ee^{-r/(2M)} =16M^3\left[(\dot{y}^2-\dot{\xi}^2)\frac{\partial r}{\partial \xi}-2\dot{\xi}\dot{y}\frac{\partial r}{\partial y}\right]\frac{\rm d}{{\rm d}r}\left(\frac{ \ee^{-r/(2M)}}{r}\right)
\nonumber \\
&+8M^2\dot{\phi}^2\xi \ee^{-r/(2M)}.
\label{Lagrange-eq-xi-Euclidean-1}
\end{align}
%\[\frac{\tilde{M}\ddot{\xi}\ee^{-r/(2M)}}{r}=\frac{\tilde{M}}{2}(\dot{y}^2-\dot{\xi}^2)\frac{\partial r}{\partial\xi}\frac{\partial}{\partial r}\left(\frac\e{e^{-r/(2M)}}{r}\right)+8M^2\dot{\phi}^2\xi \ee^{-r/(2M)},\]
The above equation can be readily simplified. Indeed, to compute $\frac{\partial r}{\partial\xi}$ we differentiate the relation \eqref{zeta-y-squared-relation} and  arrive at 
\begin{align}
2\xi=\ee^{r/(2M)}\frac{\partial}{\partial\xi}\left(\frac{r}{2M}-1\right)+\left(\frac{r}{2M}-1\right)\frac{\partial(\ee^{r/(2M)})}{\partial\xi},
%=\frac{r\ee^{r/(2M)}}{4M^2}\frac{\partial r}{\partial\xi},    
\end{align}
which  yields the identity
\begin{align}
\frac{\partial r}{\partial\xi}=\frac{8M^2\xi \ee^{-r/(2M)}}{r},  
\label{factor-Lagrange-1}
\end{align}
%\[\frac{{\rm d}}{{\rm d}\lambda}\left(\frac{\partial\mathcal{L}}{\partial\dot{\xi}}\right)=\frac{{\rm d}}{{\rm d}\lambda}\left(\frac{\tilde{M}\dot{\xi}\ee^{-r/(2M)}}{r}\right)=\frac{\tilde{M}\ddot{\xi}e^{-r/(2M)}}{r}+\tilde{M}\dot{\xi}\left(\dot{\xi}\frac{\partial r}{\partial\xi}+\dot{y}\frac{\partial r}{\partial y}\right)\frac{\rm d}{{\rm d}r}\left(\frac{\ee^{-r/(2M)}}{r}\right).\]
the expression for $\frac{\partial r}{\partial y}$ being easily obtained via the substitution $\xi \to y$;  moreover, we can work out the derivative $\frac{\rm d}{{\rm d}r}\left(\frac{\ee^{-r/(2M)}}{r}\right)$  by exploiting once again  Eq. \eqref{zeta-y-squared-relation}, which permits to find 
\begin{align}
\frac{\rm d}{{\rm d}r}\left(\frac{\ee^{-r/(2M)}}{r}\right)=     -\frac{1}{r^2}\left(\frac{r}{2M}+1\right)\ee^{-r/(2M)}= -\left(\frac{\frac{1}{4M^2}-\frac{1}{r^2}}{\xi^2+y^2}\right).
\label{factor-Lagrange-2}
\end{align}
%\[\frac{\rm d}{{\rm d}r}\left(\frac{\ee^{-r/(2M)}}{r}\right)=-\frac{\frac{{\rm d}(r\ee^{r/(2M)})}{{\rm d}r}}{(r\ee^{r/(2M)})^2}=-\frac{\ee^{r/(2M)}(r/(2M)+1)}{(r\ee^{r/(2M)})^2}=-\frac{1}{r^2}\left(\frac{r}{2M}+1\right)\ee^{-r/(2M)}\]
%further reduces to
%\[\ddot{\xi}=\frac{4M^2}{r^2}(\xi(\dot{\xi}^2-\dot{y}^2)+2y\dot{\xi}\dot{y})\left(\frac{r}{2M}+1\right)\ee^{-r/(2M)}+\frac{J^2\xi}{4Mr^3}.\]%=\frac{\xi}{4Mr^3}\left(\frac{\tilde{M}r}{2}(\dot{\xi}^2-\dot{y}^2)\left(\frac{r}{2M}+1\right)\ee^{-r/(2M)}+J^2\right)\]
%and  using \eqref{zeta-y-squared-relation} to work out $\frac{\frac{r}{2M}+1}{r^2}\ee^{-r/(2M)}$ 
%\[\frac{\frac{r}{2M}+1}{r^2}\ee^{-r/(2M)}=\frac{\frac{1}{4M^2}-\frac{1}{r^2}}{\xi^2+y^2},\]
Thus, it follows from Eqs. \eqref{factor-Lagrange-1} and \eqref{factor-Lagrange-2} that  Eq. \eqref{Lagrange-eq-xi-Euclidean-1} boils down to
\begin{align}
\ddot{\xi}=\frac{\xi(\dot{\xi}^2-\dot{y}^2)+2y\dot{y}\dot{\xi}}{\xi^2+y^2}\left(1-\frac{4M^2}{r^2}\right)+\frac{J^2\xi}{4Mr^3}.    
\end{align}
Therefore, the geodesic equations written in Kruskal-Szekeres coordinates in the equatorial plane of the Euclidean Schwarzschild spacetime read as
\begin{subequations}
\label{geo_sys}
\begin{align}
\dot{\phi}&=\frac{J}{r^2}, 
\label{geo_sys-1}
\\
\ddot{\xi}&=\frac{\xi(\dot{\xi}^2-\dot{y}^2)+2y\dot{y}\dot{\xi}}{\xi^2+y^2}\left(1-\frac{4M^2}{r^2}\right)+\frac{J^2\xi}{4Mr^3}, 
\label{geo_sys-2}
\\
\ddot{y}&=\frac{y(\dot{y}^2-\dot{\xi}^2)+2\xi\dot{\xi}\dot{y}}{\xi^2+y^2}\left(1-\frac{4M^2}{r^2}\right)+\frac{J^2y}{4Mr^3}.
\label{geo_sys-3}
\end{align}
\end{subequations}

\subsubsection{Proof of the geodesic completeness }\label{Sec:geodesic-completeness}

We are now ready to prove  that the real Riemannian section of the complexified Schwarzschild spacetime is geodesically complete. 

Without loss of generality, we now let the affine parameter $\lambda\equiv s$, $s$ being the  arc length. This means that the tangent vector $\dot{X}^\mu$ to the geodesic $X^\mu (s)$ is normalized, and we identically have, on the equatorial plane, 
\begin{equation}\label{normalized}
\mathcal{G}_{\mu\nu}\dot{X}^\mu\dot{X}^\nu=\frac{32M^3\ee^{-r/(2M)}}{r}(\dot{\xi}^2+\dot{y}^2)+r^2\dot{\phi^2}\equiv1,
\end{equation}
where we recall that all the coordinates take real values in the Euclidean domain.

We let the map $s \mapsto X^\mu(s)$ be maximally defined on an open interval $I=]s_-,s_+[$ containing $0$, and we aim to prove that $s_\pm=\pm\infty$. This amounts to show that the solutions of the system \eqref{geo_sys} are globally defined. However, since the geodesic equation 
%\[\ddot{\gamma}^\alpha+{\Gamma^\alpha}_{\mu\nu}\dot{\gamma}^\mu\dot{\gamma}^\nu=0\]
is invariant under the exchange transformation $s\leftrightarrow -s$, we only need to demonstrate that $s_+=+\infty$. To establish our result, we will assume for contradiction that this fails, i.e.,  we suppose that  
\begin{equation}\label{assumption} 
\begin{array}{l}
\text{the function $X^\mu(s)$ is  defined only for $s \in I = ]s_-,s_+[$ with  $s_+<+\infty$.}  
\end{array}
\end{equation}
Then, we will see that this non-globality condition leads  to an absurdity via the \textit{explosion alternative theorem}. In this way,   we will be able to conclude that our claim is verified: the interval $I$  really coincides with the whole real line $\mathbb{R}$. Recall that the explosion alternative theorem describes the behaviour of a solution of a differential equation   as the limits of its domain are approached. In our settings, since $s\mapsto X^\mu(s)\in\R^2\times\mathbb{S}^1$ is supposed to be maximal on $I$ and $s_+<\infty$ by hypothesis \eqref{assumption}, the explosion alternative   ensures that $X^\mu(s)$ approaches the boundary of $\R^2\times\mathbb{S}^1$ as $s\to s_+$ (equivalently, we can say that the points $X^\mu(s)$ must escape any compact subset of $\R^2\times\mathbb{S}^1$ as $s\to s_+$). In particular,  this amounts to say that $s\mapsto(\xi(s),y(s))\in\R^2$ is unbounded on any neighborhood of $s_+$ contained in $I$. For further details regarding the explosion alternative theorem, we refer the reader to paragraph 17.4 in Ref. \cite{hirsch-smale-devaney} and Corollary 3.2 in Ref. \cite{hartman-ode}.

In the forthcoming demonstration, we will need the following elementary but useful result: 
\begin{equation}\label{cauchy} 
\begin{array}{l}
\text{If $f(\tilde x)$ is a continuously differentiable function of some variable $\tilde{x}$ defined}  \\
\text{on an interval $[a,b[$ with $b<+\infty$,  and if the derivative of $f$ is bounded,} \\
\text{then  the real-valued limit $\lim \limits_{\tilde x \to b^-} f(\tilde x)$ exists and is finite. }
\end{array}
\end{equation}
This follows from the fact that such  $f$ is a  Lipschitz function  on $[a,b[$, by the mean value theorem, together with the Cauchy criterion for functions \cite{mahmudov}.

We know from Eq. \eqref{geo_sys-1} that the Schwarzschild geometry admits a  constant of motion $J\in\R$. The time independence of the metric implies the presence of another first integral  representing the \textit{energy}, which is defined, in Schwarzschild coordinates $x^\mu$,  as the momentum conjugate to $\tau$
\begin{align}
    E:=p_\tau =g_{\mu \nu} \left(\partial_\tau\right)^\mu \dot{x}^\nu,
\end{align}
where $\left(\partial_\tau\right)^\mu=\left(\frac{\partial}{\partial\tau}\right)^\mu$ is the static Killing vector field. Differentiating formula  \eqref{y-and-xi-relations} with respect to $\tau$ yields 
\begin{align}
\partial_\tau=\frac{\partial\xi}{\partial\tau}\partial_\xi+\frac{\partial y}{\partial\tau}\partial_y=\frac{1}{4M}(y\partial_\xi-\xi\partial_y),    
\end{align}
which leads to following expression for the energy in Kruskal--Szekeres coordinates: 
\begin{equation}\label{energy_KS}
E=\frac{1}{4M}(y \mathcal{G}_{\xi\mu}\dot{X}^\mu-\xi \mathcal{G}_{y\mu}\dot{X}^\mu)=\frac{8M^2\ee^{-r/(2M)}}{r}(y\dot{\xi}-\xi\dot{y}).
\end{equation}
We now claim that the coordinates $\xi$ and $y$ are proportional along the geodesic $X^\mu(s)$ exactly when $E=0$, i.e., 
\begin{equation}\label{E=0}
E=0~\Longleftrightarrow~y\propto\xi.
\end{equation}
First of all, from the formula \eqref{energy_KS}, we immediately have $E=0$ if and only if 
\begin{align}
y\dot{\xi}-\xi\dot{y}=0.
\label{equality-proof}
\end{align}
Moreover, observe  that the the existence and uniqueness theorem applied to Eqs. \eqref{geo_sys-2} and \eqref{geo_sys-3} shows that if the quadruple $(\xi,\dot{\xi},y,\dot{y})$ vanishes for some value of the parameter $s$, then it must vanish everywhere. Consequently,   from Eq. \eqref{zeta-y-squared-relation}, $r$ remains constant, as well as $\dot{\phi}=Jr^{-2}$, and the geodesic $X^\mu(s)$ would thus be globally defined, contrary to our initial assumption \eqref{assumption}. We can therefore suppose that $(\xi,\dot{\xi},y,\dot{y})$ does not vanish identically and hence   invoke a classical Wronskian theorem of Peano \cite{bocher}. Indeed, Eq. \eqref{equality-proof} is simply the statement that the Wronskian  $W(\xi,y)\equiv y\dot\xi-\xi\dot y $
of the functions $\xi$ and $y$ vanishes. This in turn implies, by the aforementioned Peano theorem,    that  $\xi$ and $y$ are linearly dependent, hence the claim \eqref{E=0}.

We will now demonstrate that the assumption $s_+<+\infty$ implies $E=0$, and thus $y\propto\xi$ by Eq.  \eqref{E=0}. If we choose the initial condition $(\xi(0),y(0))=(0,0)$, then Eq. \eqref{energy_KS} readily gives $E=0$, while from Eq. \eqref{zeta-y-squared-relation} we  obtain $r(0)=2M$, which remains constant owing to Eq. \eqref{radial_motion} below. Then, we select $(\xi(0),y(0))\neq (0,0)$ so that we may assume $r(0)>2M$. Then, combining Eqs. \eqref{geo_sys-2} and \eqref{geo_sys-3} with Eq.  \eqref{zeta-y-squared-relation}, we  arrive at the radial geodesic equation (see Ref. \cite{Battista-Esposito2022} for details)
\begin{equation}\label{radial_motion}
\dot{r}^2=\left(1-\frac{2M}{r}\right)\left(1-\frac{J^2}{r^2}\right)-E^2.
\end{equation}
Taking the derivative of the above equation, we arrive at the radial Euler--Lagrange equation, which can be  recast as the following first-order differential equation 
\begin{subequations}
\label{initial-value-problem}
\begin{align}
\dot{R}=f(R),
\end{align}
\text{with   initial condition }
\begin{align}
R(0)=(r(0),\dot{r}(0)),    
\end{align}
\end{subequations}
where we have introduced the shorthand notation $R:=(r,\dot r)$, and $f: ]2M,+\infty[\times\R\to\R^2$ is defined by
\[f\left(\rho,\dot{\rho}\right):=\left(\dot{\rho},\frac{M\rho^2+J^2\rho-3MJ^2}{\rho^{4}}\right).\]

Equation \eqref{initial-value-problem} clearly represents a well-posed initial value problem, thereby allowing the application of the existence and uniqueness theorem. Thus, by the explosion alternative theorem, since $(r,\dot{r})\in]2M;+\infty[\times\R$ is a maximal solution of the  problem \eqref{initial-value-problem}  on some sub-interval $I'=]s'_-,s'_+[\subset I$,  then the map $s \mapsto (r(s),\dot{r}(s))$ approaches the boundary of $]2M;+\infty[\times\R$ as $s\to s_+'$. This means that, when $s\to s_+'$, the function $s\mapsto r(s)$ must come arbitrarily close to $2M$, or to $+\infty$, or both.  However, since $r$ is bounded from below in the Euclidean section (see Eq.\eqref{r-bigger-2M}), Eq. \eqref{radial_motion} implies that $\dot{r}^2$ is bounded on $I'$,  and since $s'_+\le s_+<+\infty$, the criterion \eqref{cauchy} then shows that $r(s)$ admits a finite limit as $s<s'_+$ approaches $s'_+$, and this limit can be nothing but $2M$. Therefore,  taking the limit of the right-hand side of Eq. \eqref{radial_motion} as $s\to s'_+$ yields 
\begin{align}\label{lim-cond}
0\le\lim_{\substack{s\to s'_+ \\ s<s'_+}}\left[\left(1-\frac{2M}{r(s)}\right)\left(1-\frac{J^2}{r(s)^2}\right)-E^2\right]=\left(1-\frac{2M}{2M}\right)\left(1-\frac{J^2}{4M^2}\right)-E^2=-E^2.
\end{align}
Therefore, it follows from Eq. \eqref{radial_motion} that $\lim\limits_{\substack{s\to s'_+ \\ s<s'_+}}\dot{r}^2=-E^2$, and hence in order to have a real-valued radial velocity we must conclude that $E=0$.  

Having just proved that $E=0$ if $s_+<+\infty$ , then from Eq. \eqref{E=0} we know that functions $y$ and $\xi$ are linearly dependent. However, the non-globality condition \eqref{assumption}  implies that they cannot both vanish identically (otherwise they would  be defined on the whole real line $\mathbb{R}$), and since the situation is symmetric in $(\xi,y)$, we may assume $\xi\ne0$. Therefore, we can find a constant $\alpha\in\R$ such that $y=\alpha\xi$ so that the normalization condition \eqref{normalized} becomes
\begin{align}
\dot{\xi}^2=\frac{r\ee^{r/(2M)}}{32M^3(1+\alpha^2)}\left(1-\frac{J^2}{r^2}\right).    
\end{align}
%and inserting this in the second equation from \eqref{geo_sys} \[\ddot{\xi}=\frac{2M(1+\alpha^2)\xi\dot{\xi}^2e^{-r/(2M)}}{r}\left(1+\frac{2M}{r}\right)+\frac{L^2\xi}{4Mr^3}\] leads to the equation \[\ddot{\xi}=\frac{\xi}{4Mr}\left(1-\frac{1}{2}\left(1-\frac{r}{2M}\right)\left(1-\frac{L^2}{r^2}\right)\right).\]
%\[\ddot{\xi}=\frac{\xi}{16M^2}\left(1+\frac{2M}{r}\right)\left(1-\frac{L^2}{r^2}\right)+\frac{L^2\xi}{4Mr^3}\]
%\[\ddot{\xi}=\frac{\xi}{4Mr}\left(\frac{r}{4M}\left(1+\frac{2M}{r}\right)\left(1-\frac{L^2}{r^2}\right)+\frac{L^2}{r^2}\right)\]
%\[\ddot{\xi}=\frac{\xi}{4Mr}\left(\frac{1}{2}\left(1+\frac{r}{2M}\right)\left(1-\frac{L^2}{r^2}\right)+\frac{L^2}{r^2}\right)\]
Now, as we know that the variable $r$ cannot blow-up at a finite parameter, the right-hand side of the above equality is bounded on $[0,s_+[$, and so is $\dot{\xi}$. The sought-after contradiction then follows from a final application of the criterion \eqref{cauchy}. Specifically, since we have just shown that $\dot{\xi}$ is bounded on $[0,s_+[$,  it follows from Eq. \eqref{cauchy} then $\xi$ has a finite limit as $s\to s_+$. Consequently,   $y$ also has a finite limit, but  this conclusion contradicts the explosion alternative theorem, which, on the contrary, predicts that either $\xi$ or $y$ be unbounded as  $s$ approaches  $s_+$.  Therefore, the hypothesis \eqref{assumption} cannot be sustained and this proves the geodesic completeness of the   Euclidean section of the Schwarzschild spacetime, as we must conclude that $X^\mu(s) $ is defined in $I \equiv \mathbb{R}$. In particular, it is worth mentioning that Eq. \eqref{radial_motion} entails that any geodesic with $J=E=0$ and negative initial radial velocity will eventually hit the hypersurface $r=2M$  within a finite amount of the affine parameter, and then go to infinity.

Thanks to the geodesic completeness, the Euclidean Schwarzschild spacetime $\mathcal{M}$ acquires nice topological and metric properties.  Indeed, since the metric   is Riemannian, the Hopf--Rinow theorem applies and implies that the compact subsets of $\mathcal{M}$ are precisely those which are both closed and bounded. In particular, $\mathcal{M}$ is complete as a metric space. Moreover, any two points in $\mathcal{M}$ can be joined by a length-extremizing geodesic. These features are  in radical contrast with the  Lorentzian-signature Schwarzschild  solution \cite{Wald-book}.

Some final remarks are  in order. First of all, observe that in the conventional Lorentzian framework the sign of the energy appearing in the radial equation \eqref{radial_motion} is reversed, and thus a condition of the form \eqref{lim-cond} no longer implies the vanishing of $E$. This is one of the main reasons why the above arguments fail in the Lorentzian world. Moreover, our proof has been performed by considering the Kruskal--Szekeres extension \eqref{Schwarzschild-metric-2} of the Schwarzschild metric and heavily relies on the fact that the variables $\xi$ and $y$ are real in the Euclidean section. The variable $\tau$ and its underlying periodicity have thus played no role in our demonstration, a fact that ties in with the Schwarzschild geometry being static.

\subsection{The complex Schwarzschild metric} \label{Sec:Complex-Schwarzschild-1}

Both  the Euclidean-signature and Lorentzian-signature Schwarzschild metrics can be described by resorting to the concept of sections of the complexified spacetime, which can be outlined as follows. If we let the variable $\xi$  be complex,  then the subspace where $\xi$ is purely imaginary corresponds to the Lorentzian section, where $\zeta$ is real  (cf. Eq. \eqref{xi-and-z}) and the metric attains the Lorentzian-signature expression  \eqref{Lorentzian-2}. Conversely,  the subspace where $\xi$ is real pertains to the  Euclidean section, where the metric is given by Eq. \eqref{Schwarzschild-metric-2}. As pointed out before, in this domain  both $\xi$ and $y$ are real, while  $\zeta$ is imaginary. Similarly, we can define a complex $\tau$-plane. Then, when  $\tau$ is   purely imaginary, we recover the Lorentzian domain, where $t$ becomes real (see Eq. \eqref{tau-and-t}), while for $\tau$ real (and periodic) we have the Euclidean section, with   the metric assuming the form \eqref{Schwarzschild-metric-1}. Therefore, the  metric is  real on both  the Lorentzian and Euclidean sections,  but it is generally complex if $\xi$ or $\tau$ are neither real nor purely imaginary. 

We now consider the Euclidean Schwarzschild metric in its most general form, by allowing for values of $\tau$ not belonging to the Euclidean domain. Therefore,  we hereafter regard $\tau$ as a complex  variable,  while $r$, $\theta$, and $\phi$ remain real-valued. We express $\tau$  as 
\begin{align}
\tau =  T + \ii t, \qquad  (T \in \mathbb{R}, t \in \mathbb{R}),
\label{complex-tau-variable}
\end{align}
and  regard
\begin{align}
 \dd s^2 =    \left(1-\frac{2M}{ r}\right) \dd \tau^2  +  \left(1-\frac{2M}{ r }\right)^{-1} \dd r^2  + r^2 \left({\rm d}\theta^2+\sin^2\theta \, {\rm d} \phi^2\right),
 \label{Schwarzschild-new-1}
\end{align}
as a complex-valued Schwarzschild metric. The time component reads explicitly as $\dd \tau^2= \left(\dd T + \ii \dd t \right)^2 = \dd T^2 - \dd t^2 + 2 \ii \dd t \dd T$. Thus,  in line with the preceding arguments, the metric boils down to the real-valued Lorentzian-signature Schwarzschild solution \eqref{Lorentzian-1} when $T=0$; on the other hand,
 when $t=0$, then $\tau$ is real (and periodic) and we recover the Euclidean section, where the metric \eqref{Schwarzschild-new-1} assumes real values (cf. Eq. \eqref{Schwarzschild-metric-1}).

The presence of the complex  variable \eqref{complex-tau-variable} introduces  one further real dimension, which means that the spacetime manifold has now five real dimensions. The coordinate transformations which bring Eq. \eqref{Schwarzschild-new-1} to the Kruskal-Szekeres form can be easily worked out as follows (see e.g. the formulas reported in  Appendix A of Ref. \cite{Capozziello2024}). If we introduce the complex-valued variables
\begin{align}
u^\prime &= \ee^{-\ii u/(4M)}, 
 \nonumber \\
v^\prime &= \ee^{ \ii v/(4M)}, 
\label{variable-u-v-prime}
\end{align}
where
\begin{align}
u&= \tau +\ii \, r^\star, 
 \nonumber \\
v&= \tau - \ii \, r^\star, 
\label{variables-u-and-v}
\end{align}
 the tortoise coordinate $r^\star$ being defined as usual \cite{Poisson2009,Carroll2004}
\begin{align}
r^\star = \int \frac{\dd r}{1- 2M/r}= r + 2M \log \left(\frac{r}{2M} -1 \right), 
\label{tortoise-coordinate} 
\end{align}
then the metric \eqref{Schwarzschild-new-1} becomes
\begin{align} 
\dd s^2 &= \left(1-\frac{2M}{r}\right) \dd u \, \dd v  +r^2 \left({\rm d}\theta^2+\sin^2\theta \, {\rm d}\, \phi^2\right) 
\nonumber \\
&= \frac{32M^3}{r} \ee^{-r/(2M)}  \, \dd u^\prime \, \dd v^\prime +r^2 \left({\rm d}\theta^2+\sin^2\theta \, {\rm d}\, \phi^2\right). \label{Schwarzschild-new-2}
\end{align}
Here, the variable $r$ is given implicitly  from 
\begin{align}
u^\prime v^\prime =   \left( \frac{r}{2M}-1\right) \ee^{r/(2M)}, 
\label{prodict-u-and-v-prime}
\end{align}
which yields 
\begin{align}
  r= 2M \left[ 1+ \mathcal{W} \left(u^\prime v^\prime / \ee\right)\right],  
\end{align}
$\mathcal{W}$ being the Lambert function \cite{Corless1996,Valluri2000}. 

As pointed out before, outside the Euclidean domain  $\xi$ and $y$ become complex. Therefore,  at this stage, let us we define (cf. Eq. \eqref{y-and-xi-relations})
\begin{align}
\xi &= -\frac{\ii}{2} \left(v^\prime - u^\prime\right),    
\nonumber \\
y &= \frac{1}{2} \left(v^\prime + u^\prime\right),   
\end{align}
then Eq. \eqref{Schwarzschild-new-2} yields (cf. Eq. \eqref{Schwarzschild-metric-2})
\begin{align}
\dd s^2 =  \frac{32M^3}{r} \ee^{-r/(2M)}  \left( \dd \xi^2 + \dd y^2 \right)  +r^2 \left({\rm d}\theta^2+\sin^2\theta \, {\rm d}\, \phi^2\right),
\label{Schwarzschild-new-3}
\end{align}
where $r$ is given by (cf. Eqs. \eqref{y-and-xi-relations} and  \eqref{zeta-y-squared-relation})
\begin{align}
r= 2M \left[ 1+ \mathcal{W} \left(\frac{\xi^2 + y^2}{\ee} \right) \right],  
\label{r-Lambert}
\end{align}
with  
\begin{align} \label{xi-y-squared-new}
\xi^2 + y^2 = u^\prime v^\prime.  
\end{align}
It follows from the above relations that, as long as $r$ is real, both the product $ u^\prime v^\prime $  and, as a consequence, the combination $\xi^2 + y^2$ will be real
(see Eq. \eqref{prodict-u-and-v-prime}).   Therefore,  the complex variables $u^\prime $ and $v^\prime$, as well as  $\xi$ and $y$, can be written in terms of three rather than four real-valued variables. This is consistent with the fact that the number of real dimensions of the complexified spacetime amounts to five.  However, in these settings  Eq. \eqref{xi-y-squared-new} does not guarantee the condition \eqref{r-bigger-2M}. Indeed, if one expresses the complex $\xi$ and $y$ as 
\begin{align}
\xi&= \xi_\Re + \ii \xi_\Im, 
\nonumber \\
y &=y_\Re + \ii y_\Im, 
\label{xi-y-complex}
\end{align}
respectively,  then the real-valued sum $\xi^2 + y^2$ gives
\begin{align}
\xi^2 + y^2 = \left(y_\Im^2+\xi_\Im^2\right)\left(y_\Re^2-\xi_\Im^2\right)/ \xi_\Im^2,
\label{xi-2-y-2-sum}
\end{align}
which is not  positive definite. The consequences of  this will become clear  in Sec. \ref{Sec:Geodesic-equations}, where we will see that the $r=0$ singularity can be attained by following some  properly generalized geodesics.

For future purposes, it is convenient to introduce coordinates $x^a =(T,t,r,\theta,\phi)$ (with the Latin indices spanning five dimensions and ranging as $a,b, \dots=0,1,2,3,4$). In this way, the five-dimensional complex-valued metric  \eqref{Schwarzschild-new-1} can be written explicitly as 
\begin{align}
\dd s^2&=g_{ab}\,\dd x^a\,\dd x^b
\nonumber \\
&=\left(1-\frac{2M}{r}\right)(\dd T^2-\dd t^2+2\ii\,\dd T\,\dd t)+\left(1-\frac{2M}{r}\right)^{-1}\dd r^2+r^2\left({\rm d}\theta^2+\sin^2\theta \, {\rm d}\, \phi^2\right).   
\label{Schwarzschild-complex-explicit}
\end{align}
This metric is degenerate, as its determinant vanishes identically. Specifically,  the eigenvalues are  $[0,0,r^2,r^2\sin^2\theta,(1-2M/r)^{-1}]$,    which means that  the degeneracy is confined to the temporal components of $g_{ab}$. We can explain this scenario   as follows. For the non-degenerate, real-valued Lorentzian and Euclidean Schwarzschild solutions, it is meaningful to speak of  an arrow of time because the time variable resides on the real line, which has a natural order relation. On the other hand, a  complex time variable admits no such arrow, since the space of complex numbers $\C$ lacks any inherent order. Consequently, there is no \qm{preferred direction} for time in the complex case, a situation that accounts for the temporal part of the metric becoming degenerate. These considerations apply not only to the complexified Schwarzschild geometry discussed here,  which, as pointed out before, represents our prototype model, but also to other solutions we will explore in Sec. \ref{Sec:applications-S-RN-FLRW}. Therefore,  metrics with vanishing temporal eigenvalues  are the most suited for describing time-complexified spacetimes, which we  might refer to as  ``arrowless spacetimes''.

Degenerate metrics open up numerous intriguing possibilities and applications. As outlined in the introduction, they have been extensively investigated in classical cosmological and black-hole geometries (see e.g. Refs. \cite{stoica-singular,Searight2017,Gunther,Klinkhamer-Wang2019,Battista2020,Wang2021,Holdom2023a,Capozziello2024}), whereas in quantum scenarios they are mainly related to space topology changes in the spacetime, a mechanism which is expected to be a key feature in quantum gravity \cite{Horowitz1990,Louko1995}. In addition, first-order gravity involving Hilbert-Palatini action comprises both invertible and non-invertible tetrad configurations, where the former yield a description equivalent to the standard second-order formulation, while  the latter enable solutions with a nonzero torsion even in absence of  matter \cite{Kaul2016,Kaul2016b,Kaul2018}. This point has some significant consequences in  quantum gravity in the first-order form, where both invertible and non-invertible tetrad fields should be taken into account when performing functional integrations \cite{Kaul2016,Kaul2016b,Tseytlin1981}. Furthermore, Ashtekar formalism \cite{Ashtekar1987}, which is essential for the development of loop quantum gravity \cite{Ashtekar2021}, remains well defined also for degenerate real triads, which in fact lead to a particular degenerate extension of general relativity \cite{Jacobson1987,Jacobson1996}.

Contrary to many of the aforementioned frameworks, where the metric becomes degenerate on a  zero-measure set,  in our settings the complex Schwarzschild metric \eqref{Schwarzschild-complex-explicit} is degenerate in the whole spacetime. In spite of this issue, we propose to construct a theory where the ill-defined inverse metric tensor is replaced by a pseudoinverse metric, which extends the notion of  inverse  metric tensor. This task will be performed by providing a covariant generalization of the  well-known Moore-Penrose algorithm \cite{Moore1920,Penrose1955}, which allows us to construct sensible curvature tensors. This result guarantees that the ensuing equations  hold independently of  the coordinates adopted, as in conventional differential geometry.  The development of this new method will be pursued in Sec. \ref{Sec:generalized-Moore-Penrose-algorithm}.

\section{Generalized covariant Moore-Penrose algorithm} \label{Sec:generalized-Moore-Penrose-algorithm}

The Moore-Penrose algorithm was first devised by Moore and later refined by Penrose  to extend the process of matrix inversion to cases where traditional inversion is not possible \cite{Moore1920,Penrose1955}.  This construction is widely used in both  pure mathematics and theoretical physics  due to its ability to  facilitate various analytical and computational applications \cite{Arfken1995,Strang2014,Baksalary2021}. 

Specifically, given a generic matrix $\mathbb{A}$, there exists a unique matrix $\mathbb{A}^+$, called the Moore-Penrose inverse of $\mathbb{A}$ or pseudoinverse, satisfying the following properties \cite{Penrose1955,Barata2012}: 
\begin{subequations}
\label{Moore-Penrose-conditions}
\begin{align}
\mathbb{A} \mathbb{A}^+ \mathbb{A} &= \mathbb{A},
\label{MP-1} \\
\mathbb{A}^+ \mathbb{A} \mathbb{A}^+ &= \mathbb{A}^+,
\label{MP-2}\\
\left(\mathbb{A} \mathbb{A}^+\right)^* &= \mathbb{A} \mathbb{A}^+, 
\label{MP-3}\\
\left(\mathbb{A}^+ \mathbb{A}\right)^* &= \mathbb{A}^+ \mathbb{A}.
\label{MP-4}
\end{align}
\end{subequations}
It is  clear from the above conditions that  $\mathbb{A}^+$ boils down to the standard inverse $\mathbb{A}^{-1}$  if $\mathbb{A}$ is invertible; moreover, when   only the requirement \eqref{MP-1} is fulfilled,  $\mathbb{A}^+$ is referred to as generalized inverse and is not  unique in general, since uniqueness demands the full set of relations \eqref{Moore-Penrose-conditions} to be met.

In this section, we devise an  extension of the Moore-Penrose procedure  that can be applied to any  symmetric degenerate tensor having constant (but non-maximal) rank.  We will refer to this method as the \textit{covariant Moore-Penrose algorithm}. This technique permits to work out a pseudoinverse metric satisfying a set of  \textit{covariant} identities, which  generalize the standard formulas \eqref{Moore-Penrose-conditions} and are thus valid independently of the particular coordinate system chosen (see Sec. \ref{Sec:pseudoinverse-metric}). In this approach, a crucial role is played by the adjoint operation, which is covariantly evaluated via a fixed real-valued metric tensor $\zeta_{ab}$. By adopting this novel strategy,  it is possible to extend  the usual concept of covariant derivative by introducing a  connection that,  in general, includes torsion (see Sec. \ref{Sec:covariant-derivative-operator}). Moreover, as demonstrated in Sec. \ref{Sec:torsion-and-curvature},  in our  new framework  the Riemann tensor constructed by means of the metric and its covariant pseudoinverse is well-defined, ensuring that all quantities derived from it behave like tensors (as we will see,  these objects represent true tensors because no terms involving the derivatives of the pseudoinverse metric occur in their expressions). We conclude the section by providing an  application to the Schwarzschild geometry and other well-known solutions in  Sec. \ref{Sec:applications-S-RN-FLRW}.

\subsection{The pseudoinverse metric}\label{Sec:pseudoinverse-metric}

As pointed out before,  the standard Moore-Penrose inverse of a degenerate matrix is uniquely defined using conditions \eqref{Moore-Penrose-conditions}. While  properties  \eqref{MP-1} and \eqref{MP-2} constitute covariant relations, special care is required  when dealing with  Eqs. \eqref{MP-3} and \eqref{MP-4}, since  the ordinary adjoint of matrices \textit{depends} on the choice of coordinates. Therefore,  to develop  the  covariant Moore-Penrose algorithm, it is essential to first 
 provide a covariant prescription for the adjoint. This task can be performed  by introducing a fixed base Riemannian metric $\zeta_{ab}$ on the spacetime manifold. The choice of this  metric is, in a sense, the price to pay to obtain a well-defined tensorial Moore-Penrose inverse. We now make this setup more precise.

First of all, we recall that in our setup the metric is represented by a  degenerate complex-valued   $(2,0)$-tensor field $g_{ab}$ which can be formally seen  as a global section of $T^*\mathcal{M}\otimes T^*\mathcal{M}$, where $T^*\mathcal{M}$ is the cotangent bundle of $\mathcal{M}$. Now, let $(\mathcal{M},\zeta)$ be a connected Riemannian manifold.  The metric $\zeta$ permits to define the following adjoint operations
\begin{align}\label{right-dagger}
\cdot^\dagger : {X_a}^b \longmapsto  \zeta_{ac}{\overline{X}_d}^c\zeta^{db} \equiv \left({X_a}^b\right)^\dagger,
\end{align}
and, dually, 
\begin{align}\label{left-dagger}
    {}^\dagger{\cdot}  : {Y^a}_b  \longmapsto  \zeta^{ac}{\overline{Y}^d}_c\zeta_{db} \equiv {}^\dagger\left({Y^a}_b\right),
\end{align}
the overbar denoting the complex conjugation. 
In practice, we have defined a ``conjugate transpose'' of tensors that is \qm{twisted} via the metric $\zeta_{ab}$. This application represents the covariant version for the standard complex adjoint of a matrix, and shares some  similarities with the usual formula for the adjoint of a matrix with respect to some inner  product.

The adjoint constructions \eqref{right-dagger} and \eqref{left-dagger} can be  formally identified as the (complex) \textit{adjoint maps}  between spaces of tensor fields
\begin{align}
&\cdot^\dagger :  T^*_\C\mathcal{M} \otimes T_\C\mathcal{M} \longmapsto   T^*_\C\mathcal{M}\otimes T_\C\mathcal{M},
\label{right-dagger-spaces}
\\
&{}^\dagger{\cdot}  :  T_\C\mathcal{M} \otimes T^*_\C\mathcal{M} \longmapsto  T_\C\mathcal{M}\otimes T^*_\C\mathcal{M},
\label{left-dagger-spaces}
\end{align}
which might be  dubbed the right and left dagger maps, respectively.  In the above equations,  we have introduced the complexified spaces  $T_\C\mathcal{M}$ and $T^*_\C\mathcal{M}$, which are essential  in order for the  complex conjugation operation performed in Eqs. \eqref{right-dagger} and \eqref{left-dagger}  to be well-defined. The former is defined by $T_\C\mathcal{M}:=T\mathcal{M}\otimes\C$, and  represents a  tangent bundle whose fibers are complexified as vector spaces. This means  that  a complex tangent vector is expressed as $X=u^a\partial_a+\ii v^a\partial_a$, with $u^a,v^a$ smooth real functions. Following a similar procedure, we have  $T^*_\C\mathcal{M}:=T^*\mathcal{M}\otimes\C$\footnote{Some care has to be taken: for $p\in\mathcal{M}$, the space $(T^*_\C)_p\mathcal{M}=T_p^*\mathcal{M}\otimes\C$ is not the complex dual space of $(T_\C)_p\mathcal{M}=T_p\mathcal{M}\otimes\C$. It rather is the complexification of the real dual space.}, which represents a complexified  cotangent bundle. The maps \eqref{right-dagger-spaces} and \eqref{left-dagger-spaces}  can be expressed by employing an intrinsic language which does not resort to coordinates, as demonstrated in Appendix \ref{Sec:technical-stuff} (see Eq. \eqref{adjoints}).

The covariant dagger relations \eqref{right-dagger}  and \eqref{left-dagger} are crucial for  the  construction of a generalized covariant Moore-Penrose pseudoinverse metric $\tilde{g}$. Indeed, given a degenerate metric $g$,  we define $\tilde{g}$ as the unique $(0,2)$-tensor field (formally, $\tilde g$ represents a section of $T_\C\mathcal{M}$) satisfying the following conditions: 
\begin{subequations}
\label{Moore-Penrose-metrici}
\begin{align}
g\tilde{g}g &= g,
\label{property-pseudo-1i}
\\
\tilde{g}g\tilde{g} &= \tilde{g},
\label{property-pseudo-2i}
\\
\left(g\tilde{g}\right)^\dagger &= g\tilde{g},
\label{property-pseudo-3i}
\\
{}^\dagger{\left(\tilde{g}g\right)} &= \tilde{g}g,
\label{property-pseudo-4i}
\end{align}    
\end{subequations}
where each product is understood as a contracted tensor product of fields. Using Eqs. \eqref{right-dagger} and \eqref{left-dagger}, we can thus recast Eq. \eqref{Moore-Penrose-metrici} in  components
\begin{subequations}
\label{Moore-Penrose-metricc}
\begin{align}
g_{ac}\tilde{g}^{cd}g_{db} &= g_{ab},
\label{property-pseudo-1c}
\\
\tilde{g}^{ac}g_{cd}\tilde{g}^{db} &= \tilde{g}^{ab},
\label{property-pseudo-2c}
\\
\zeta^{bd}{{g_{de}\tilde{g}^{ec}}}\zeta_{ca} &= \overline{g_{ac}\tilde{g}^{cb}},
\label{property-pseudo-3c}
\\
\zeta_{bd}{\tilde{g}^{de}g_{ec}}\zeta^{ca} &= \overline{\tilde{g}^{ac}g_{cb}}.
\label{property-pseudo-4c}
\end{align}    
\end{subequations}
The use of the previously defined covariant adjoint operation guarantees that
the full set of conditions \eqref{Moore-Penrose-metrici}, or equivalently \eqref{Moore-Penrose-metricc}, is independent of the coordinate system adopted, in stark contrast to the conventional Moore-Penrose relations  \eqref{Moore-Penrose-conditions}. As a consequence, we can claim that $\tilde{g}^{ab}$ is a well-defined tensor field on the spacetime manifold $\mathcal{M}$.

A crucial point is now the proof of the existence and uniqueness of  $\tilde{g}$, which  can be established by means of  the following argument.  First of all,  let us consider  the particular case where $\mathcal{M}=U\subset\R^n$ is an open subset of $\R^n$ and $\zeta=\delta_{ij}\dd x^i\otimes\dd x^j$ ($i,j=1,\dots,n$)  the usual Euclidean metric. Since by hypothesis $g$ has constant rank, its usual Moore-Penrose inverse is a smooth tensor on $U$. Slightly more generally, if $\zeta$ is any Riemannian metric on $U$, then its Cholesky decomposition\footnote{\label{footnote-1}Given a symmetric positive-definite matrix $\mathbb{A}$,  there exists a unique upper-triangular matrix $\mathbb{B}$ with positive diagonal that provides the so-called \textit{Cholesky decomposition}  $\mathbb{A}={}^t\mathbb{B}\mathbb{B}$. The matrix $\mathbb{B}$ is called the \textit{Cholesky factor} of $\mathbb{A}$.} $\zeta={}^tBB$ permits to construct the \qm{translated} tensor $g_0={}^t{B^{-1}}gB^{-1}$, which has a classical pseudoinverse $\tilde{g}_0$. Thus,  the \qm{retranslated} tensor $\tilde{g}:=B^{-1}\tilde{g}_0{}^t{B^{-1}}$ is a pseudoinverse for $g$ with respect to $\zeta$.  In other words, the Cholesky factor of $\zeta$ furnishes a change of coordinates  that is used to \qm{pull back} the usual Moore-Penrose inverse  of the coordinate-transformed $g_0$ of $g$. The smoothness of the Cholesky factor implies that the result is a tensor on $U$, which is uniquely defined because the usual Moore-Penrose inverse is. The full generality is then achieved by a classical argument: we take an atlas on $\mathcal{M}$ to locally construct a pseudoinverse and the uniqueness statement ensures that these local tensors agree on overlapping charts, so that they glue together into a well-defined tensor, finishing the proof.  Observe moreover that the uniqueness condition entails that $\tilde{g}$ is symmetric, since $g$ is symmetric. The details of this demonstration are given in Appendix \ref{Sec:technical-stuff}.

The advantage of  adopting our covariant inversion mechanism lies in the fact it allows us to can claim that the procedure of raising the indices of a tensor via  $\tilde{g}^{ab}$ is  a well-defined and unambiguous operation, since the pseudoinverse metric is uniquely defined by the set of covariant identities \eqref{Moore-Penrose-metricc}.   This result is crucial for the definition of the covariant derivative operator and  curvature tensors, which will be investigated in the next sections.

As a final remark, it is worth stressing that our algorithm is valid for both real-valued and complex-valued degenerate metrics. In particular, the method can be adopted to compute the covariant pseudoinverse of  any symmetric degenerate tensor, as  the above recipe can be  easily generalized  to  handle  tensors of any rank (in this regard, 
it is worth noting that a pseudoinversion mechanism has been  proposed in Ref. \cite{atindogbe-ezin-tossa}, but it is limited only to certain degenerate hypersurfaces of codimension 2; moreover,   the Moore-Penrose inversion of linear tensors has been introduced in Ref. \cite{liang}). Furthermore, we  notice that  the fixed metric $\zeta_{ab}$ assumes  the Riemannian signature  $(+++++)$, as it is required for evaluating the Cholesky decomposition (as well as its smoothness, see  formula \eqref{eq:smooth_cholesky}).

\subsection{Covariant derivative operator and torsion tensor} \label{Sec:covariant-derivative-operator}

Now that our covariant inversion method has allowed us to obtain a uniquely defined symmetric tensor $\tilde{g}^{ab}$, we can   investigate whether the spacetime manifold $\mathcal{M}$ admits an affine connection and a  covariant derivative operator.

Let us start with the well-known formula for the first-kind Christoffel symbols \cite{Wald-book} 
\begin{align}\label{def-christoffel}
    \Gamma_{cab}:=\frac12\left(\partial_ag_{bc}+\partial_bg_{ac}-\partial_cg_{ab}\right).
\end{align}
Remarkably,  these are the coefficients of the so-called Sasin--Eledrisi \qm{contravariant connection}, which we will consider in Sec. \ref{Sec:Extremal-curves}. A natural idea to define a connection on $\mathcal{M}$ would be to mimic the definition of the usual second-kind Christoffel symbols via the expression $\tilde{g}^{dc}\Gamma_{cab}$. However, while the resulting connection would  be torsion-free, it would not preserve the metric in general. For this reason, we  consider a  derivative operator $\nabla$  whose action on the basis vectors $\{\partial_a\}$ of  $T_\C\mathcal{M}$  reads as 
\begin{align}\label{def-nabla}
 \nabla_{\partial_a } \left(\partial_b\right) \equiv \nabla_a \left(\partial_b\right) \equiv  {\Gamma^d}_{ab} \partial_d,
\end{align}
so that  the following (non-symmetric)  connection coefficients can be derived:
%{\begin{align}
%    {\Gamma^c}_{ab} := \tilde{g}^{cd}(\Gamma_{dab}+\Lambda_{bad})=\tilde{g}^{cd}(\partial_ag_{bd}-g_{be}\tilde{g}^{ef}\Gamma_{fad}),
%\end{align}}
\begin{align}\label{second-christoffel}
    {\Gamma^a}_{bc}:=\tilde{g}^{ad}(\Gamma_{dbc}+\Lambda_{cbd})+{{\hat\Gamma^a}_{bc}}-\tilde{g}^{ad}g_{de}{\hat\Gamma^e}_{bc}.%(\zeta^{mn}-\tilde{g}^{mu}g_{uv}\zeta^{vn}){}^\zeta{\Gamma}_{nkj}.
\end{align}
Intuitively, this  represents the simplest $g$-invariant connection involving the ``naive Christoffel symbols'' $\tilde{g}^{dc}\Gamma_{cab}$; see Appendix \ref{Sec:unique-connection} for a precise statement. In Eq. \eqref{second-christoffel},  ${\hat\Gamma^d}_{ab}$ denotes the Levi-Civita connection associated to the Riemannian metric $\zeta_{ab}$, and  we have introduced the  ``raising-lowering'' tensor $\Lambda_{abc}$
\begin{align}\label{def-Lambda}
    \Lambda_{abc}:=({\delta_a}^d-g_{ae}\tilde{g}^{ed})\Gamma_{dbc}=\Gamma_{abc}-g_{ad}\tilde{g}^{de}\Gamma_{ebc},
\end{align}
which clearly satisfies
\begin{align}\label{gl=0}
    \tilde{g}^{ad}\Lambda_{dbc}=0,
\end{align}
in view of Eq. \eqref{property-pseudo-2c}. We have checked that   $ {\Gamma^a}_{bc}$  satisfies the ordinary transformation law for connections under a change of coordinates. This result guarantees that the covariant derivative of a tensor  transforms covariantly   under general coordinate transformations. In this regard, it is worth noticing that any combination of the type $\gamma - g \tilde{g} \gamma$, $\gamma$ being a generic (even non-symmetric) connection, yields a genuine affine connection. In particular, the Bianchi identities, which we will derive in Sec. \ref{Sec:torsion-and-curvature} below, always assume the same functional form when a symmetric $\gamma$ is adopted. Therefore, the combination ${{\hat\Gamma^a}_{bc}}-\tilde{g}^{ad}g_{de}{\hat\Gamma^e}_{bc}$
occurring in the formula \eqref{second-christoffel} appears to be the simplest and most natural choice for the construction of an affine connection. 

The tensor $\Lambda_{abc}$ fulfills a crucial role in our analysis. In fact, it clearly vanishes  whenever  $g_{ab}$ has maximal rank, whereas it is in general not zero for degenerate metrics, as in this case  $g_{ae} \tilde{g}^{ed} $ is no longer the identity. However, there exist degenerate metrics with a high degree of symmetry for which the combination \eqref{def-Lambda} yields $\Lambda_{abc}=0$, such as the Euclidean Schwarzschild  and  Reissner-Nordstr\"{o}m solutions, which will be studied in  Sec. \ref{Sec:applications-S-RN-FLRW}.

Having constructed a true covariant derivative, it is easy to see that, for any complex vector fields $X,Y,Z$ defined on $\mathcal{M}$ (and intended as sections of $T_\C\mathcal{M}$), the operator $\nabla$ satisfies the usual relations \cite{Wald-book,Nakahara2003} 
\begin{subequations}
\label{nabla-properties}
\begin{align}
&\nabla_X (Y+Z)= \nabla_X Y+ \nabla_X Z,
\\
&\nabla_{X+Y} Z=\nabla_X Z + \nabla_Y Z,
\\
&\nabla_{fX}Y=f\nabla_X Y, 
 \\
&\nabla_X(fY)=X(f) Y+f\nabla_XY,
\end{align}
\end{subequations}
$f$ being a complex-valued smooth function.  Therefore, $\nabla$ acts on any vector $V=V^a \partial_a$ as 
\begin{align}
\nabla_a V^b= \partial_a V^b +  {\Gamma^b}_{ad}  V^d.
%\nonumber \\
%=&\hat{\nabla}_aV^b+V^d\left[\tilde{g}^{bc}(\Gamma_{cad}+\Lambda_{dac})-\tilde{g}^{be}g_{ef}{\hat\Gamma^f}_{ad}\right],
 %\nabla_a V^b= \partial_a V^b + V^d \left(\Gamma_{cad}+\Lambda_{dac}\right) \tilde{g}^{cb}.
 \label{covariant-derivative-1}
\end{align}
 
As we have proved in Appendix \ref{Sec:unique-connection}, the connection \eqref{second-christoffel} is   $g$-invariant, i.e., 
\begin{align}\label{g-invariance}
\nabla_a g_{bc}=0.
\end{align}
On the other hand, the covariant derivative of the covariant  pseudoinverse metric  is in general not zero, as
\begin{align}
\nabla_a\tilde{g}^{bc}=&\partial_a\tilde{g}^{bc}+\left[\tilde{g}^{bd}(\Gamma_{dae}+\Lambda_{ead})+{\hat\Gamma^b}_{ae}-\tilde{g}^{bd}g_{df}{\hat\Gamma^f}_{ae}\right]\tilde{g}^{ec} 
    \nonumber \\
&+\left[\tilde{g}^{cd}(\Gamma_{dae}+\Lambda_{ead})+{\hat\Gamma^{c}}_{ae}-\tilde{g}^{cd}g_{df}{\hat\Gamma^f}_{ae}\right]\tilde{g}^{be} 
\nonumber\\
 %=&\partial_a\tilde{g}^{bc}+\tilde{g}^{bd}\tilde{g}^{ce}\partial_ag_{de}+({\hat\Gamma^b}_{ae}-\tilde{g}^{bd}g_{df}{\hat\Gamma^f}_{ae})\tilde{g}^{ec}+({\hat\Gamma^{c}}_{ae}-\tilde{g}^{cd}g_{df}{\hat\Gamma^f}_{ae})\tilde{g}^{be} \\
=&\partial_a\tilde{g}^{bc}+\tilde{g}^{bd}\tilde{g}^{ce}\partial_ag_{de}+2\left(\tilde{g}^{e(c}{\hat\Gamma^{b)}}_{ae}-\tilde{g}^{e(c}\tilde{g}^{b)d}g_{df}{\hat\Gamma^f}_{ae}\right)\ne0.
\label{covariant-tilde-g}
\end{align}

The affine connection \eqref{second-christoffel} has a nonvanishing torsion tensor $ T(X,Y):=\nabla_XY-\nabla_YX-[X,Y]$, whose components read as
\begin{align}\label{twisted-torsion-tensor-coefs}
    {T_{ab}}^c={\Gamma^c}_{ab}-{\Gamma^c}_{ba}=(\Lambda_{bau}-\Lambda_{abu})\tilde{g}^{uc}.
\end{align}
Our covariant prescription implies that this object is a true  tensor. Remarkably, in our framework ${T_{ab}}^c$ is not sourced by  the presence of some spinning matter fields.  Therefore, we have found the interesting result that the effects of having a non-invertible metric can be geometrically mapped  into the presence of a torsion tensor, which vanishes for  metrics that have maximal rank or  highly symmetric degenerate geometries.  This conclusion aligns with  the findings of Refs. \cite{Kaul2016,Kaul2016b}, which show that first-order gravity with degenerate tetrads allows for solutions with  a nonvanishing torsion tensor of purely geometric origin.

\subsection{Riemann curvature tensor}\label{Sec:torsion-and-curvature}

One of the advantages of our covariant Moore-Penrose algorithm is that it permits  to define a genuine Riemann tensor $R(X,Y):=[\nabla_X,\nabla_Y]-\nabla_{[X,Y]}$, with  components
\begin{align}\label{contra-riemann-coefs}
    {R^a}_{bcd}=\langle \dd x^a, (\nabla_c\nabla_d-\nabla_d\nabla_c)\partial_b  \rangle=\partial_c{\Gamma^a}_{db}-\partial_d{\Gamma^a}_{cb}+{\Gamma^a}_{ce}{\Gamma^e}_{db}-{\Gamma^a}_{de}{\Gamma^e}_{cb}, 
\end{align}
$\langle \cdot , \cdot\rangle$ being the ordinary inner product. The $g$-invariance \eqref{g-invariance} implies the conventional  symmetries 
\begin{align}\label{bianchi-symmetry}
    R_{(ab)cd}=R_{ab(cd)}=0,
\end{align} 
whereas the property $R_{abcd}=R_{cdab}$ is spoiled by the presence of ${T_{ab}}^c$ (see Eq. \eqref{property-riemann-appendix} in Appendix \ref{Sec:properties-proofs}), a result in line with the general features of Riemann-Cartan geometry  \cite{Hehl1976,Ortin2015} (see also Eq. (75) in Ref. \cite{Battista2021}). The  torsion tensor also affects the first and second Bianchi identities, which take the generalized forms:   
\begin{align}
  &  {R^a}_{[bcd]}=\nabla_{[b}{T_{cd]}}^a-{T_{[bc}}^e{T_{d]e}}^a, %\tilde{g}^{au}g_{uv}\nabla_{[b}{T_{cd]}}^v+{T_{u[b}}^a{T_{cd]}}^u,
\label{bianchi-I} \\
&\nabla_{[e}{R^a}_{|b|cd]}+{R^a}_{bu[e}{T_{cd]}}^u=0, %{R^a}_{b[cd;e]}+{R^a}_{bu[c}{T_{de]}}^u=0,
\label{bianchi-II}
\end{align}
in agreement with standard Riemann-Cartan formulas  \cite{Penrose-Rindler-1,Ortin2015},  the corresponding Riemannian relations being recovered in the torsion-free case (we refer again to   Appendix \ref{Sec:properties-proofs} for a detailed calculation).

The Ricci tensor $R_{ab}$ stems from the non-trivial contraction of the Riemann tensor, which we perform via the covariant pseudoinverse according to the formula
\begin{align}
R_{ab}&:=\tilde{g}^{cd}R_{cbda} \equiv {R^c}_{bca},
\label{Ricci-def-tilde-g}
\end{align}
which gives
\begin{align}
R_{ab}=  \partial_c{\Gamma^c}_{ab}-\partial_a{\Gamma^c}_{cb}+{\Gamma^c}_{cd}{\Gamma^d}_{ab}-{\Gamma^c}_{ad}{\Gamma^d}_{cb}.
  \label{ricci-coefs}
\end{align}
Taking the trace of the first Bianchi identity \eqref{bianchi-I}, we find that the antisymmetric part of   $R_{ab}$ satisfies the relation 
\begin{align}\label{comm-ricci}
    R_{[ab]}=\frac12(\nabla_{[a}{T_{cb]}}^c-{T_{[ac}}^d{T_{b]d}}^c),
\end{align}
which shows that the Ricci tensor is in general not symmetric, like in conventional Riemann-Cartan spacetime \cite{Ortin2015,Medina2018}. By computing the trace of $R_{ab}$ via the pseudoinverse metric, we obtain
\begin{align}\label{ricci-scalar}
    R:=\tilde{g}^{ab}R_{ab}. 
\end{align}

Thanks to the generalized Moore-Penrose scheme, the Ricci tensor \eqref{ricci-coefs}, the Ricci scalar \eqref{ricci-scalar}, and, as a consequence, the Einstein tensor
\begin{align}
G_{ab}:=R_{ab}-\frac12 R g_{ab}, 
\end{align}
represent well-defined quantities. This key result is guaranteed by the fact that, on the one hand, the pseudoinverse metric $\tilde{g}^{ab}$ is a true covariant tensor, and, on the other, no derivatives of $\tilde{g}^{ab}$ appear in these curvature-related objects.

The contracted Bianchi identity attains a generalized form in our setup, owing to both the nonzero torsion tensor and the fact that the covariant pseudoinverse metric is, in general, not covariantly conserved (see Eq. \eqref{covariant-tilde-g}). Starting from formula \eqref{bianchi-II} and exploiting the identities $    \tilde{g}^{bc}{T_{ec}}^u=\tilde{g}^{bc}\tilde{g}^{uk}(\Lambda_{cek}-\Lambda_{eck})=-\tilde{g}^{bc}\tilde{g}^{uk}\Lambda_{eck}$ and   $\tilde{g}^{bc}\tilde{g}^{ad}{T_{cd}}^u=0$, which follow from Eqs. \eqref{gl=0} and \eqref{twisted-torsion-tensor-coefs},   we obtain, after some calculation,  
\begin{align}\label{contracted-bianchi}
\nabla_eR=2\left[\nabla^aR_{ea}+R_{bc}\nabla_e\tilde{g}^{bc}+R_{abce}\nabla^a\tilde{g}^{bc}+R_{ab}\tilde{g}^{bc}{T_{ce}}^a\right],
\end{align}
or, equivalently, 
\begin{align}\label{contracted-bianchi-bis}
    \nabla_eR=2\left[\nabla^aR_{ea}+R_{bc}\nabla_e\tilde{g}^{bc}+R_{abce}\nabla^a\tilde{g}^{bc}-\tilde{g}^{ab}R_{bc}\tilde{g}^{cd}\Lambda_{eda}\right],
\end{align}
which boil down to the Riemannian identity $\nabla_eR=2\nabla^aR_{ae}$  when  $\nabla_a\tilde{g}^{bc}=0$ and ${T_{ab}}^c$, or equivalently $\Lambda_{abc}$, vanish.

In the scenarios where  $\Lambda_{abc}=0$,  the additional symmetry $R_{cdab}=R_{abcd}$,  jointly with the property \eqref{bianchi-symmetry},  allows us to raise and then lower all four indices in the Riemann tensor, as in conventional geometries. For example, for the first index,  we have $R_{abcd}=g_{ae}\tilde{g}^{ef}R_{fbcd}$. Therefore, it follows from Eq. \eqref{contracted-bianchi-bis} with $\Lambda_{abc}=0$ that  the conservation law for the Einstein tensor takes the modified form 
\begin{align}\label{conservation-1}
    \nabla^aG_{ab}=\left(2R_{a[c}g_{d]b}-R_{cdab}\right)\nabla^c\tilde{g}^{da},~~~~\text{(when $\Lambda_{abc}=0$),}%\nabla^aG_{ab}=\left(g_{bv}R_{uw}-g_{bu}R_{vw}-R_{uvwb}\right)\nabla^u\tilde{g}^{vw}~~~\text{(when $\Lambda=0$).}
\end{align}
where we have taken into account that the Ricci tensor is symmetric in this case. 
Remarkably, in the even more special situation where $\nabla_a\tilde{g}^{bc}=0$, we retrieve the classical Einstein conservation law 
\begin{align}\label{conservation-law}
    \nabla^a G_{ab}=0,~~~~\text{(when $\Lambda_{abc}=\nabla_a\tilde{g}^{bc}=0$).}
\end{align}

\subsection{Some applications}\label{Sec:applications-S-RN-FLRW}

The analysis of the previous section  has suggested the existence of a \qm{special} class of degenerate geometries where all the  formulas derived so far simplify significantly. These models are set apart by the validity of the following relations:
\begin{subequations}
\begin{align}
\Lambda_{abc}&=0,
\label{Schwrazschild-simplif-cond-1}
\\
{\hat\Gamma^a}_{bc}-\tilde{g}^{ad}g_{de}{\hat\Gamma^e}_{bc}&=0,
\label{Schwrazschild-simplif-cond-2}
\\
\nabla_a \tilde{g}^{bc}&=0.
\label{Schwrazschild-simplif-cond-3}
\end{align}
\label{standard-conditions}
\end{subequations}
The first identity alone is equivalent to the vanishing of the torsion tensor (see Eq. \eqref{twisted-torsion-tensor-coefs}), while  the two properties  \eqref{Schwrazschild-simplif-cond-1} and \eqref{Schwrazschild-simplif-cond-2} imply that the connection coefficients \eqref{second-christoffel}  attain the Riemannian-like form 
\begin{align}
{\Gamma^a}_{bc}=\tilde{g}^{ad} \Gamma_{dbc}. 
\label{connection-symmetric-Schwarzschild}
\end{align}
This simplified version of the affine connection  yields the following expression for the Riemann tensor   \eqref{contra-riemann-coefs}: 
\begin{align}\label{simplified-Riemann}
R_{abcd}=\partial_c\Gamma_{adb}-\partial_d\Gamma_{acb}+\tilde{g}^{ef}\left(\Gamma_{ecb}\Gamma_{fda}-\Gamma_{edb}\Gamma_{fca}\right),
\end{align}
which thus resembles the Riemannian formula valid for nondegenerate metrics.  Moreover, as a consequence of Eqs. \eqref{Schwrazschild-simplif-cond-1} and \eqref{Schwrazschild-simplif-cond-3}, the Einstein tensor satisfies the ordinary conservation law \eqref{conservation-law}.  

The scope of this  section is twofold. On the one hand, we provide some applications of the covariant Moore-Penrose algorithm, thereby demonstrating the existence of the aforementioned \qm{special} frameworks.  Specifically,  we analyze  the Schwarzschild geometry in Sec.   \ref{Sec:application-Schwarzschild}, and the Reissner--Nordstr\"{o}m and Reissner--Nordstr\"{o}m-de Sitter solutions in Sec.   \ref{Sec:application-Reissner}. For these complexified black-hole spacetimes, the relations   \eqref{standard-conditions} are always respected. On the other hand, for the cosmological Friedmann--Lema\^{i}tre--Robertson--Walker (FLRW) paradigm investigated in Sec. \ref{Sec:application-FLRW}, we will see that the $\Lambda_{abc}$ tensor is nonvanishing under certain conditions.

\subsubsection{Schwarzschild geometry}\label{Sec:application-Schwarzschild}

As pointed out before,   the complex  Schwarzschild metric \eqref{Schwarzschild-complex-explicit} is degenerate, as its  determinant vanishes identically. Explicitly, the metric tensor $g_{ab}$ can be written through the following   matrix:  
\begin{align}
g_{ab}=
\begin{pmatrix}
1-2M/r & \ii(1-2M/r) & 0 & 0 & 0 \\ \ii(1-2M/r) & -(1-2M/r) & 0 & 0 & 0 \\ 0 & 0 & (1-2M/r)^{-1} & 0 & 0 \\ 0 & 0 & 0 & r^2 & 0 \\ 0 & 0 & 0 & 0 & r^2\sin^2\theta
\end{pmatrix}.
\label{Schwarzschild-metric-matrix-form}
\end{align}
For this geometry, the base metric $\zeta_{ab}$ occurring in the right and left dagger maps  \eqref{right-dagger} and \eqref{left-dagger} amounts to the five-dimensional flat Euclidean metric. In coordinates $x^a=(T,t,r,\theta,\phi)$, it is  given by
\begin{align}
\zeta_{ab}\,\dd x^a\,\dd x^b=\vert \dd \tau \vert^2 +\dd r^2+r^2 \left(\dd\theta^2+\sin^2\theta\,\dd\phi^2 \right)=\dd T^2+\dd t^2+\dd r^2+r^2 \left(\dd\theta^2+\sin^2\theta\,\dd\phi^2 \right),
\label{zeta-Schwarzschild}
\end{align}
and admits a diagonal Cholesky factor  $B={\rm diag}\left(1,1,1,r,r \sin \theta \right)$.  Since $g_{ab}$  commutes with $\zeta_{ab}$,  it also commutes with its Cholesky factor $B$. Therefore, the usual Moore-Penrose inverse of $g_{ab}$ commutes with $B$ as well. This implies that the covariant pseudoinverse $\tilde{g}^{ab}$ satisfying the identities \eqref{Moore-Penrose-metricc}  boils down to the standard Moore-Penrose inverse of $g_{ab}$, and is thus given by 
\begin{align}
\tilde{g}^{ab}=\begin{pmatrix}
\frac{1}{4 \left(1-2M/r\right)} & -\frac{\ii}{4\left(1-2M/r\right)} & 0 & 0 & 0\\
-\frac{\ii}{4\left(1-2M/r\right)} & -\frac{1}{4 \left(1-2M/r\right)} & 0 & 0 & 0 \\
0 & 0 & 1-2M/r & 0 & 0 \\
0 & 0 & 0 & r^{-2} & 0 \\
0 & 0 & 0 & 0 & r^{-2} \sin^{-2} \theta
\end{pmatrix}.
\label{pseudoinverse-metric}
\end{align}
Therefore, thanks to the covariant inversion prescription developed in Sec. \ref{Sec:pseudoinverse-metric}, we can claim that $\tilde{g}^{ab}$ is a well-defined tensor, although it can be calculated via the standard Moore-Penrose construction. Indeed, our calculations demonstrate that the ordinary Moore-Penrose scheme represents a particular case of a more general covariant framework, at least for the  Schwarzschild model.

Remarkably, the complex Schwarzschild geometry satisfies the relations \eqref{standard-conditions}. As a consequence, the connection is given by formula \eqref{connection-symmetric-Schwarzschild}, which, contrary to the general case, is compatible with the metric.  The Ricci tensor \eqref{Ricci-def-tilde-g}  can be easily constructed through the simplified expression
\eqref{simplified-Riemann}, and it is found to be zero, like for the real-valued solutions pertaining to  the Euclidean and Lorentzian sections of the complexified spacetime (see Sec. \ref{Sec:complexified-Schwarzschild-geom}). This means that the conservation law \eqref{conservation-law} is trivially  satisfied, as should be expected from the fact that  Eqs. \eqref{Schwrazschild-simplif-cond-1} and \eqref{Schwrazschild-simplif-cond-3} hold.

Starting from Eq. \eqref{simplified-Riemann}, we have determined  that the  Kretschmann invariant assumes the same value as in the standard case, i.e.,
\begin{align}
R_{abcd} R^{abcd}= \frac{48M^2}{r^6},
\label{Kretschmann-Schwarzschild}
\end{align}
a result which holds in any coordinate system owing to  the application of the covariant Moore-Penrose formalism. This allows us to conclude that the only curvature singularity characterizing the complexified Schwarzschild spacetime is located at $r=0$, like for the ordinary solution.

Some  crucial remarks are now in order. The results of this section prove that the Schwarzschild model represents such a special geometry that it renders all the subtleties of the formalism developed in this paper  not evident.  Indeed, formula \eqref{connection-symmetric-Schwarzschild} might induce one to infer that a torsion-free connection of the type $\tilde{g}^{ad} \Gamma_{dbc}$ is the correct one to use when dealing with degenerate metrics. However, as we have described in the previous sections, this is generally not true. 

Since the metric \eqref{Schwarzschild-metric-matrix-form}  satisfies the vacuum  Einstein equations $R_{ab}=0$,  it can be interpreted  as a  generalized gravitational instanton. The main differences with the ordinary solution are its degenerate character and the fact that it is singular, as  from our proof of Sec. \ref{Sec:geodesic-completeness}  we know that only the Euclidean domain of the complexified spacetime is geodesically complete. For this new class of instantons,  an extended notion of geodesics will be  provided  in Sec. \ref{Sec:Geodesic-equations}.

\subsubsection{Reissner--Nordstr\"{o}m and Reissner--Nordstr\"{o}m-de Sitter metrics} \label{Sec:application-Reissner}

The real-valued  maximal-rank Euclidean Reissner-Nordstr\"{o}m solution (along with the underlying particle dynamics) has been investigated in detail in Ref. \cite{Garnier2024}\footnote{See e.g. Refs. \cite{Pugliese2010,Rahimov2024hol,Viththani2024} for some recent studies of the geodesic
motion framed in the Lorentzian-signature Reissner-Nordstr\"{o}m model and some extensions
thereof.}. By considering its complex extension following a route similar to the Schwarzschild case, which in particular makes use of Eq. \eqref{complex-tau-variable}, we find that also for this geometry the metric $g_{ab}$ is degenerate in its temporal components, the eigenvalues being $[0,0,r^2,r^2\sin^2\theta,(1-2M/r+Q^2/r^2)^{-1}]$; in addition, it is easy to show that the relations \eqref{standard-conditions} hold once again,  where for the evaluation of $\tilde{g}^{ab}$ we have considered the same   $\zeta_{ab}$ as for the Schwarzschild spacetime (see Eq. \eqref{zeta-Schwarzschild}). Furthermore, the ordinary Einstein equations are valid, as we are going to show. First of all, we adopt again coordinates $x^a =(T,t,r,\theta,\phi)$ and let 
\begin{align}
A=A_b\dd x^b=-\ii \frac{Q}{r}(\dd T+\ii \dd t),
\label{EM-one-form}
\end{align}
be the electromagnetic (one-form) potential, $Q$ being the (real-valued) electric charge. Then, the (degenerate) ``electromagnetic tensor'' $F_{ab}$ is defined by 
\begin{align}\label{expression-fab}
    F_{ab}=\partial_b A_{a}-\partial_a A_{b}=\dfrac{Q}{r^{2}}\left(\begin{smallmatrix}0 & 0 & \ii & 0 & 0 \\ 0 & 0 & -1 & 0 & 0 \\ -\ii & 1 & 0 & 0 & 0 \\ 0 & 0 & 0 & 0 & 0 \\ 0 & 0 & 0 & 0 & 0\end{smallmatrix}\right),
\end{align}
with $F^{ab}$ given by 
\begin{align}
F^{ab}=\tilde{g}^{ac}F_{cd}\tilde{g}^{db}=\frac{Q}{2r^2}\left(\begin{smallmatrix}0 & 0 & \ii & 0 & 0 \\ 0 & 0 & 1 & 0 & 0 \\ -\ii & -1 & 0 & 0 & 0 \\ 0 & 0 & 0 & 0 & 0 \\ 0 & 0 & 0 & 0 & 0\end{smallmatrix}\right).
\end{align}
In this way,  as in the  Lorentzian-signature Reissner--Nordstr\"{o}m solution, we obtain
\begin{align}\label{trace-fab}
    F^{ab}F_{ab}=-\frac{2Q^2}{r^4}, 
\end{align}
and, in addition,  the Maxwell equations
\begin{align}\label{cons-law-fab}
    \nabla_aF^{ab}=0,
\end{align}
are verified. Moreover, as in the conventional case, we have $\tilde{g}^{ab}R_{ab}=0$, and hence mimicking the classical definition of the stress-energy tensor (in units where $\mu_0=4\pi$),
\begin{align}\label{se-tensor}
    \mathcal{T}_{ab}=\frac{1}{4\pi}\left(\tilde{g}^{cd}F_{ca}F_{db}-\tfrac14g_{ab}F^{cd}F_{cd}\right), 
\end{align}
we have checked that the ordinary Einstein equations
\begin{align}
G_{ab}=8\pi \mathcal{T}_{ab},
\end{align}
are satisfied. Moreover,  the conservation law \eqref{conservation-law} is respected. 

It is worth  mentioning  that the Kretschmann scalar has the same form as for the classical Reissner--Nordstr\"{o}m spacetime, its expression being given by
\begin{align}
R_{abcd}R^{abcd}=\dfrac{8}{r^{8}}(6M^2r^2-12MrQ^2+7Q^4). 
\label{Kretschmann-RN}
\end{align}

We can introduce a cosmological constant $\Lambda\in\R$ in our analysis without harm, thus obtaining the complex degenerate Reissner-Nordstr\"{o}m-de Sitter metric
\begin{subequations}
\begin{align}
    \dd s^2=    g_{ab}\dd x^a\dd x^b=f(r)(\dd T+i\dd t)^2+\frac{\dd r^2}{f(r)}+r^2\dd\Omega^2,
\end{align}
\text{with}
\begin{align}
f(r):=1-\frac{2M}{r}+\frac{Q^2}{r^2}-\frac{\Lambda r^2}{3},
\end{align}
\end{subequations}
and $\dd\Omega^2=\dd\theta^2+\sin^2\theta\,\dd\phi^2$. For this geometry, the identities  \eqref{standard-conditions} along with the ensuing formula \eqref{conservation-law} still hold, and the relations \eqref{EM-one-form}--\eqref{se-tensor} remain unchanged.  Last, the Einstein equations with a cosmological constant
\begin{align}
G_{ab}+\Lambda g_{ab}=8\pi \mathcal{T}_{ab}
\label{Einstein-eqs-Lambda}
\end{align}
are verified.

\subsubsection{ Friedmann--Lema\^{i}tre--Robertson--Walker geometry}\label{Sec:application-FLRW}

We now consider a complex cosmological model based on the FLRW geometry. Unlike  the spacetimes discussed before, this solution admits, under some hypotheses, a nonvanishing  $\Lambda_{abc}$ tensor. 

The   five-dimensional complex degenerate FLRW metric expressed in comoving coordinates reads as
\begin{align}
\dd s^2 = g_{ab} \dd x^a \dd x^b =(\dd T+\ii\dd t)^2+a^2(T,t)\left(\frac{\dd r^2}{1-kr^2}+r^2\dd\Omega^2\right),
\label{FLRW-complex}
\end{align}
where  $T$ and $t$  are the real and imaginary part of the variable $\tau$, defined again as in Eq. \eqref{complex-tau-variable},  $a$  the complex cosmological scale factor, and  $k\in\{-1,0,1\}$ describes the spatial  curvature.

In this setup, we can  work out the pseudoinverse metric $\tilde{g}^{ab}$ by means of the covariant procedure described before and by   defining the metric $\zeta_{ab}$ as in Eq. \eqref{zeta-Schwarzschild}. In this way, we find that  Eqs. \eqref{Schwrazschild-simplif-cond-2} and \eqref{Schwrazschild-simplif-cond-3} are still valid. However, as anticipated before, the condition \eqref{Schwrazschild-simplif-cond-1} does not hold in general, since the only nonvanishing coefficients of the tensor $\Lambda_{bcd}$ have $b\in\{T,t\}$ and $c=d$  (i.e., $\Lambda_{r\bigcdot\bigcdot}=\Lambda_{\theta\bigcdot\bigcdot}=\Lambda_{\phi\bigcdot\bigcdot}=0$, while  $\Lambda_{T\bigcdot\bigcdot}$ and $\Lambda_{t\bigcdot\bigcdot}$ are diagonal). Remarkably, upon  writing $a=a_\Re+\ii a_\Im$ in real and imaginary parts, these components  vanish exactly when the  Cauchy--Riemann conditions are satisfied:
\begin{align}
    \left\{\begin{array}{rcl}\partial_Ta_\Re & = & \partial_ta_\Im, \\ \partial_ta_\Re & = & -\partial_Ta_\Im.\end{array}\right.
\end{align}
In other words, $\Lambda_{bcd}=0$ if and only if $ a(T,t)$ is a holomorphic function of the complex variable $\tau=T+\ii t$.

Interesting implications arise when the scale factor $a(\tau)$ is holomorphic, an assumption we henceforth adopt. First of all, we find that the Ricci scalar is given by
\begin{align}
\tilde{g}^{ab}R_{ab}=6\left[-\frac{{a}''}{a}-\left(\frac{{a}'}{a}\right)^2+\frac{k}{a^2}\right],
\label{FLRW-scalar-R}
\end{align}
where, by a little abuse of notation, the complex derivatives of $a$ have been denoted with primes (i.e., ${a}'\equiv \dd a(\tau)/\dd \tau$,  and so on), as in the real case.  Therefore, the Einstein tensor reads as
\begin{align}\label{flrw-einstein}
    G_{ab}=\begin{pmatrix}\frac{3({a'}^2-k)}{a^2} & \frac{3\ii({a'}^2-k)}{a^2} & 0 & 0 & 0 \\ \frac{3\ii({a'}^2-k)}{a^2} & \frac{3(-{a'}^2+k)}{a^2} & 0 & 0 & 0 \\ 0 & 0 & \frac{2a{a''}+{a'}^2-k}{1-kr^2} & 0 & 0 \\ 0 & 0 & 0 & r^2(2a{a''}+{a'}^2-k) & 0 \\ 0 & 0 & 0 & 0 & r^2\sin^2\theta(2a{a''}+{a'}^2-k)\end{pmatrix},
\end{align}
and is seen to represent the holomorphic degenerate extension of the Einstein tensor pertaining to the usual real-valued FLRW metric. Starting from Eq. \eqref{flrw-einstein}, it is easy to check that the conservation law \eqref{conservation-law} is satisfied; furthermore, we can obtain the complex analogue of the Friedmann equations. As in conventional models \cite{Hawking_Ellis_1973,Wald-book}, let us assume the ``stress-energy tensor'' of a \textit{perfect fluid}
\begin{align}\label{complex-perfect-fluid}
    \mathcal{T}_{ab}=(\rho+p)u_au_b+pg_{ab},
\end{align}
where  $\rho=\rho(\tau)$ and $p=p(\tau)$ are holomorphic functions of  $\tau$ denoting the \textit{complex density} and the \textit{complex pressure} of the fluid, respectively, and $u_a=g_{ab}u^b$, $u^b$ being the comoving fluid velocity. A natural choice for the form assumed by $u^a$ can be worked out as follows. First of all, we observe that in the Lorentzian four-dimensional FLRW cosmology, the real-valued comoving velocity satisfies $u^\mu\partial_\mu=\partial_t$ (recall that Greek indices are four-dimensional). Then,  applying a Wick rotation $t \to \ii t $ to this expression and combining the two time coordinates $T,t$ as in Eq. \eqref{complex-tau-variable},  we arrive at the following \emph{Ansatz} for the velocity field: 
\begin{align}\label{ua}
u=u^a \partial_a:=\ii \partial_\tau =\frac{\ii}{2} \partial_T+\frac12 \partial_t,%u=u^a \partial_a:=-\ii \partial_\tau =-\frac{\ii}{2} \partial_T-\frac12 \partial_t,
\end{align}
where we have employed the Wirtinger derivative  operator  $\partial_\tau=\partial_{T+\ii t}:=\tfrac12(\partial_T-\ii\partial_t)$. In this way, we find $u^a=\left(u^T,u^t,u^r,u^\theta,u^\phi\right)= \left(\ii/2,1/2,0,0,0\right)$,   while $u_a=g_{ab}u^b=(u_T,u_t,u_r,u_\theta,u_\phi)=\left(\ii,-1,0,0,0\right)$ so that $u_a \dd x^a = \ii \dd T - \dd t = \ii \dd \tau$.  These formulas agree with the relation involving the degenerate metric $u^a=\tilde{g}^{ab}u_b$, which  extends the conventional identity valid in standard frameworks.  Moreover,  we have $u^a u_a=g_{ab} u^a u^b=\tilde{g}^{ab}u_a u_b$, with   $u^au_a=-1$  as in the ordinary scenarios. Notice that if we set $t=0$ (resp. $T=0$) in Eq. \eqref{FLRW-complex}, we recover the real-valued four-dimensional FLRW metric with Euclidean (resp. Lorentzian) signature and then  $u^\mu u_\mu=1$ (resp. $u^\mu u_\mu=-1$). In this sense, we can interpret the solution \eqref{FLRW-complex} as the generalization of both the Euclidean and Lorentzian  FLRW geometries.

By means of the above expressions, the stress-energy tensor \eqref{complex-perfect-fluid}  becomes 
\begin{align}\label{flrw-stress}
    \mathcal{T}_{ab}=\begin{pmatrix}-\rho & -\ii\rho & 0 & 0 & 0 \\ -\ii\rho & \rho & 0 & 0 & 0 \\ 0 & 0 & pg_{rr} & 0 & 0 \\ 0 & 0 & 0 & pg_{\theta\theta} & 0 \\ 0 & 0 & 0 & 0 & pg_{\phi\phi}\end{pmatrix},
\end{align}
and hence, bearing in mind Eqs. \eqref{Einstein-eqs-Lambda} and  \eqref{flrw-einstein}, both  the temporal components $G_{tt}+\Lambda g_{tt}=8\pi \mathcal{T}_{tt}$ and $G_{TT}+\Lambda g_{TT}=8\pi \mathcal{T}_{TT}$ of the field equations  yield the first Friedmann equation 
\begin{align}\label{friedmann-1}
    \frac{k-{a'}^2}{a^2}=\frac{8\pi\rho+\Lambda}{3},
\end{align}
while the second  Friedmann equation 
\begin{align}\label{friedmann-2}
    %4\pi p=\frac{\ddot{a}}{a}+\frac{\dot{a}^2-k}{2a^2}+\frac\Lambda2=\frac{\ddot{a}}{a}-\frac{8\pi\rho+\Lambda}{6}+\frac\Lambda2~\Rightarrow~
   - \frac{{a}''}{a}=-\frac{4\pi}{3}(\rho+3p)+\frac\Lambda3,
\end{align}
can be obtained in various equivalent ways, i.e.,  by combining Eq. \eqref{friedmann-1} with: (i)  $G_{rr}+\Lambda g_{rr}=8\pi \mathcal{T}_{rr}$; (ii) the angular components $G_{\theta \theta}+\Lambda g_{\theta \theta}=8\pi \mathcal{T}_{\theta \theta}$ or $G_{\phi \phi }+\Lambda g_{\phi  \phi }=8\pi \mathcal{T}_{\phi \phi }$; (iii) the trace of the field equations computed via $\tilde{g}^{ab}$. It should be noticed that the Friedmann equations \eqref{friedmann-1} and \eqref{friedmann-2} exhibit a sign flip with respect to their Lorentzian counterparts. However, this change does not apply to the contribution proportional to the parameter $k$ appearing in Eq. \eqref{friedmann-1}. This fact can be explained as follows. The  $\ii t$ term  in the complex time variable $\tau=T+\ii t$ introduces  a factor of $\ii$  in front of  any time differentiation, resulting in an overall minus sign  when evaluating the squares of first-derivative quantities or computing second-order derivatives. Since the piece involving $k$ comes with no derivative of the scale factor $a$, it remains unmodified. The same considerations also pertain to the Ricci scalar \eqref{FLRW-scalar-R}.

Defining the \textit{complex Hubble parameter} $H:={{a}'}/{a}$ and using Eq. \eqref{friedmann-1}, the second  Friedmann equation \eqref{friedmann-2} can be recast as
\begin{align}
    \rho'+3H(\rho+p)=0,
\end{align}
which we have checked to be equivalent to the mass-energy conservation 
\begin{align}\label{cons-stress}
    \nabla^a \mathcal{T}_{ab}=0.
\end{align}
This is a crucial consistency check of the validity of Eq. \eqref{conservation-law}, which we have proved to be ensured since $a$ is holomorphic. Therefore, it turns out that the conservation law \eqref{cons-stress} is \textit{equivalent} to the holomorphicity of the scale factor. This can be seen as a new physical interpretation of the notion of (nonvanishing) holomorphic function.

By setting $\rho_\Lambda:=\rho+\frac{\Lambda}{8\pi}$ and $p_\Lambda:=p-\frac{\Lambda}{8\pi}$, we arrive at the  set of holomorphic differential equations
\begin{subequations}
\begin{align}\label{hubble}
H^2 & = \frac{k}{a^2}-\frac{8\pi}{3}\rho_\Lambda, \\[.5em]
{H}'+H^2 & = \frac{4\pi}{3}(\rho_\Lambda+3p_\Lambda),
    \end{align}
\end{subequations}
which represents the complex extension of the usual Hubble system. This leads to 
the following formula for the Kretschmann scalar:
\begin{align}
R_{abcd}R^{abcd}=\frac{12}{a^4}\left[a^2{a''}^2+({a'}^2-k)^2\right]=\frac{64\pi^2}{3}\left[(\rho_\Lambda+3p_\Lambda)^2+4\rho_\Lambda^2\right], %=12({a''}^2/a^2+(H^2-k/a^2)^2)=12((H'+H^2)^2+(H^2-k/a^2)^2)=12((4\pi/3(\rho_\Lambda+3p_\Lambda))^2+(8\pi/3\rho_\Lambda)^2)=\frac{64\pi^2}{3}((\rho_\Lambda+3p_\Lambda)^2+4\rho_\Lambda^2)=\frac{64\pi^2}{3}(5\rho_\Lambda^2+6\rho_\Lambda p_\Lambda+9p_\Lambda^2).
\label{Kreetschmann-FLRW}
\end{align}
which, modulo the sign in the $k$ factor,  turns out to be formally analogous to the one holding in the Lorentzian world (see comments below Eqs. \eqref{friedmann-1} and \eqref{friedmann-2}). The Kretschmann invariant is generally complex, as it is a holomorphic function of the complex time $\tau$  (this contrasts with the black hole cases explored before, where $R_{abcd}R^{abcd}$ is  real-valued; see Eqs. \eqref{Kretschmann-Schwarzschild} and \eqref{Kretschmann-RN}), and  blows up when $a(\tau)=0$,  pointing out the occurrence of a curvature singularity.  If we relax the  assumption regarding the holomorphicity of the scale factor, then the ensuing presence of the torsion tensor might allow this  singularity  to be avoided. This process could be similar to what occurs in extended gravity models such as the Einstein-Cartan theory  \cite{Poplawski2011,Luz2019},  where, in some situations, torsion  can exert a counterbalancing effect against gravitational attraction, thereby inhibiting the development of singularities  \cite{Trautman1973,vandeVenn2024}. However, such a mechanism  comes at the cost of losing the conservation law \eqref{conservation-law} of the Einstein tensor (cf. Eq. \eqref{contracted-bianchi-bis}).

The complex cosmological paradigm studied in this section can have interesting implications in   quantum cosmology,  where the program first pioneered by Hartle and Hawking envisages the ordinary Lorentzian spacetime emerging from a Euclidean region via  a metric signature transition \cite{Hartle1983}. In this scenario, the starting point  for analyzing the evolution of the universe can be represented by the five-dimensional FLRW geometry \eqref{FLRW-complex},  and the effects of the holomorphy of the scale factor $a(\tau)$ can be further explored. Additionally,  since signature-changing metrics have been investigated also in classical cosmological settings  (see e.g. Refs. \cite{Ellis1992a,Ellis1992b,Alexandre2023a}), the devised model   might have a role   in this context as well.

\section{Extended geodesic equations}\label{Sec:Geodesic-equations}

In Secs. \ref{Sec:motion-Euclidea-domain} and \ref{Sec:geodesic-completeness}, we  analyzed the geodesics of the Euclidean Schwarzschild geometry,   proving that the Euclidean section of the complexified Schwarzschild spacetime is geodesically complete.  Since in this domain (and in the Lorentzian one as well) the metric is everywhere invertible, our investigation assumed the use of the  conventional geodesic equations.  

In this section, we provide a broader concept of geodesics suited for degenerate metrics by exploiting two different notions: autoparallel and extremal curves, which will be introduced in Secs. \ref{Sec:autoparallels} and \ref{Sec:Extremal-curves}, respectively. As we will see,  these two kinds of curves lead to   different dynamical frameworks in our settings  (cf. Eqs. \eqref{autoparallel-eq1} and \eqref{gunthers-equation} below). A concrete application to our benchmark geometry, specifically the complex Schwarzschild solution introduced earlier,  will also  be provided.

\subsection{Autoparallel curves}\label{Sec:autoparallels}

An \textit{autoparallel curve} $\gamma$ (or straightest line) is a curve along which the tangent vector is parallel-transported \cite{Nakahara2003,Carroll2004}. Let $V$ be the tangent vector to the path $x^a (\lambda)$, where $\lambda$ denotes as usual the affine parameter. Then, the autoparallel curve is described by the condition 
\begin{align}\label{autoparallel-eq0}
    \nabla_{V}V=0,
\end{align}
or equivalently 
\begin{align}
\ddot{x}^a+{\Gamma^a}_{bc}\dot{x}^b\dot{x}^c=0,
\end{align}
which, in turn, gives 
\begin{align}\label{autoparallel-eq1}
  \ddot{x}^a+\left[\tilde{g}^{ad}(\Gamma_{dbc}+\Lambda_{cbd})+{{\hat\Gamma^a}_{bc}}-\tilde{g}^{ad}g_{de}{\hat\Gamma^e}_{bc}\right]\dot{x}^b\dot{x}^c=0,
\end{align}
by expressing ${\Gamma^a}_{bc}$ via Eq. \eqref{second-christoffel} (like before, the dot stands for  the derivative with respect to the affine parameter). Notice that since the connection is metric-compatible (cf. Eq. \eqref{g-invariance}), 
%Observe that because $\nabla g=0$, we have\footnote{more generally, $\tfrac{\dd}{\dd t}g(X_{|\gamma},Y_{|\gamma})=g(\nabla_{\dot\gamma}X_{|\gamma},Y_{|\gamma})+g(X_{|\gamma},\nabla_{\dot\gamma}Y_{|\gamma})$.} 
then $\tfrac{\dd}{\dd \lambda} \left(g_{ab} \dot{x}^a \dot{x}^b\right)=2 \dot{x}^a g_{ab} \dot{x}^c \nabla_c \dot{x}^b$, and the usual conservation law holds 
\begin{align}\label{constant-lagrangian}
    \frac{\dd}{\dd \lambda}\left(g_{ab}\dot{x}^a\dot{x}^b\right)=0.
\end{align}

We can interpret Eq. \eqref{autoparallel-eq1} as a first possible extension of the conventional geodesic equation to degenerate-metric settings. In analogy with Riemann-Cartan geometry \cite{Hehl1976},   it receives a contribution from the symmetric and torsion-dependent part of the connection, while  the occurrence of the  (Levi-Civita) connection ${\hat\Gamma^a}_{bc}$ associated to the Riemannian metric $\zeta_{ab}$ comes as a distinctive consequence  of the use of the covariant Moore-Penrose formalism.

However, there exist geometries where Eq. \eqref{autoparallel-eq1} attains a simpler form. One example is furnished by the complex Schwarzschild geometry, where, as we have demonstrated before,  relations \eqref{standard-conditions} hold. As a consequence,  Eq. \eqref{autoparallel-eq1} boils down to
\begin{align}\label{autoparallel-schwarzschild}
\ddot{x}^a+\tilde{g}^{ad}\Gamma_{dbc}\dot{x}^b\dot{x}^c=0,
\end{align}
which  represents the degenerate-metric counterpart of the ordinary geodesic equation featuring general relativity. The spherical symmetry of the Schwarzschild solution allows us to limit our study of autoparallel curves to the equatorial plane $\theta= \pi/2$. In this way, from Eqs. \eqref{Schwarzschild-metric-matrix-form} and \eqref{autoparallel-schwarzschild}  we obtain the two temporal equations
\begin{subequations}
\begin{align}
\ddot{T}+\frac{M\dot{r}(\dot{T}+\ii\dot{t})}{r^2(1-2M/r)}=0, \label{for-T} \\
\ddot{t}-\ii\frac{M\dot{r}(\dot{T}+\ii\dot{t})}{r^2(1-2M/r)}=0, \label{for-t}
\end{align}
\label{temporal-schwarzschild}
\text{the radial equation}  
\begin{align}
    \ddot{r}&=\left(1-\frac{2M}{r}\right)\left\{r\dot{\phi}^2+\frac{M}{r^2}\left[\frac{\dot{r}^2}{(1-2M/r)^2}+(\dot{T}+\ii\dot{t})^2\right]\right\},
    \label{for-r-Schwarzschild-1}
\end{align}
\text{and the angular equation }
\begin{align}
\ddot{\phi}+2\dot{r}\frac{\dot{\phi}}{r}=0,
\label{for-phi}
\end{align}
\end{subequations}
the latter having  the same form as in the conventional case. By combining  Eqs. \eqref{for-T} and \eqref{for-t}, we readily obtain  $\ddot{t}+\ii\ddot{T}=0$. Since  $\gamma$ is a real curve,  the components $x^a(\lambda)$ and  their derivatives  are real functions of the affine parameter $\lambda$. This allows us to separate the real and imaginary parts,  leading to the equalities $\ddot{T}=0$ and $\ddot{t}=0$, which, in view of Eqs. \eqref{for-T} and \eqref{for-t},   either imply $\dot{r}=0$ or $\dot{T}=\dot{t}=0$. Therefore, in the Schwarzschild spacetime there exist two kinds of autoparallel curves. If $\dot{r}=0$, then  Eq. \eqref{for-phi} yields $\ddot{\phi}=0$,  and hence $\gamma$ is a circular curve with $T$ and $t$ as affine times. As in this situation both $T$ and $t$ are allowed to vary, these curves might not be entirely included in either the Lorentzian section or the Euclidean domain of the complexified Schwarzschild spacetime. On the other hand, when $\dot{T}=\dot{t}=0$, $\gamma$ is a static curve whose   angular and radial components obey the ordinary  geodesic equations of the Lorentzian-signature Schwarzschild solution. The angular equation is given by Eq. \eqref{for-phi}, whereas the radial one  follows from Eq. \eqref{for-r-Schwarzschild-1} and reads as
\begin{align}
\ddot{r}=(r-2M)\dot{\phi}^2+\frac{M\dot{r}^2}{r(r-2M)}.
%\ddot{r}+\frac{\dot{\phi}^2}{1-2M/r}\left(4M-r-\frac{4M^2}{r}\right)-\frac{M\dot{r}^2}{r^2(1-2M/r)}=0. 
\end{align}
These curves thus lie in the hypersurface where $T$ and $t$ are constant, i.e., in the intersection of the Lorentzian and Euclidean domains.

The above analysis of the Schwarzschild geometry suggests that the concept of autoparallel curves leads to a notion of geodesics which appears to be \qm{too stringent},  as this construction predicts the presence of static  curves. In the next section, we will provide a new paradigm for the definition of geodesics, which admits the possibility of letting one time component vary.

\subsection{Extremal  curves}\label{Sec:Extremal-curves}

A different  definition of geodesics arises from the notion of extremals, which correspond to paths of extremal length with respect to the metric of the manifold \cite{Hehl1976}. We can give a precise meaning to these curves also in  degenerate-metric settings by considering the ``contravariant derivative'', hereafter denoted by $\nabla^*$, first introduced  in Ref. \cite{sasin-eledrisi}. This operator has been exploited to investigate certain quantum-gravity frameworks based on noncommutative geometry  \cite{Heller1997}, and   is defined as the $\C$-linear map
\begin{align} \label{nabla-star-map}
\nabla^* : \mathcal{X}(\mathcal{M})\otimes\mathcal{X}(\mathcal{M})\to \Omega^1(\mathcal{M}),
\end{align}
where $\mathcal{X}(\mathcal{M})$ denotes the space of complex  vector fields on $\mathcal{M}$ (intended as sections of $T_\C\mathcal{M}$) and $\Omega^1(\mathcal{M})$ the space of one-forms on $\mathcal{M}$, i.e., the dual space of $\mathcal{X}(\mathcal{M})$. In other words, given two vector fields belonging to $\mathcal{X}(\mathcal{M})$, $\nabla^*$ returns a one-form instead of a new vector field (this will be evident from Eq. \eqref{nabla-star-vector} below). For this reason, and  because  it is  also    metric-compatible and torsion-free, $\nabla^*$ can be interpreted as  the ``dual'' analogue of the Levi-Civita connection. 

For any $X,Y\in\mathcal{X}(\mathcal{M})$ and for any complex-valued smooth function $f$,  $\nabla^*$ verifies the following conditions:
\begin{subequations}
\label{contra-connection}
\begin{align}
&\nabla^*_{fX}Y=f\nabla^*_XY, 
\label{contra-connection-1} \\
&\nabla^*_X(fY)=X(f)g(Y,-)+f\nabla^*_XY,
\label{contra-connection-2}
\end{align}
\end{subequations}
where  the $g$-dependent relation \eqref{contra-connection-2} replaces the usual Leibniz rule. Here, 
the notation $g(Y,-)$ indicates the one-form constructed via the map $Y^a \in T_{\C}\mathcal{M} \mapsto g_{ab} Y^b \in T^*_{\C} \mathcal{M} $.
%indicates, for any $p \in \mathcal{M}$, the one-form $ p\mapsto g(Y_p,-)$, where $g(Y_p,-)$ is the map $T_\C\mathcal{M}\ni v\mapsto g(Y_p,v)\in\C$ (equivalently, we can write $Y^\flat$ by employing  the flat canonical isomorphism linking $T_{\C}\mathcal{M}$ and $T^*_{\C} \mathcal{M}$).
The action of $\nabla^*$ on the basis vector $\partial_b$ is described by
\begin{align}\label{nabla-star-vector}
\nabla^*_{\partial_a}  (\partial_b) \equiv \nabla^*_{a}  (\partial_b) =\Gamma_{cab} \dd x^c,
\end{align}
where $\Gamma_{cab}$ are the  Christoffel symbols of the first kind defined in Eq. \eqref{def-christoffel}. As anticipated before,  the right-hand side of the above formula  features the presence of the one-form $\dd x^c$ rather than the vector field $\partial_c$, in agreement with Eq. \eqref{nabla-star-map}. Therefore, from the identities  \eqref{contra-connection} jointly with Eq. \eqref{nabla-star-vector}, we obtain the general expression
\begin{align}\label{nabla-star-general}
\nabla^*_{u^a\partial_a}(v^b\partial_b)=u^a\nabla^*_{a}(v^b\partial_b)= u^a\left[\partial_a(v^b)g_{bc}+v^b\Gamma_{cab}\right]\dd x^c.
\end{align}

Given the above premises,  we can define a ``\textit{contravariant autoparallel}'' curve  by
\begin{align}
    \nabla^*_{V}V=0,
\end{align}
where, like before, $V$ denotes the tangent to the curve. Bearing in mind Eq. \eqref{nabla-star-general}, this condition yields
\begin{align}
\left(\ddot{x}^ag_{ab}+\Gamma_{bcd}\dot{x}^c\dot{x}^d\right){\dd x^b}=0,
\end{align}
which entails that
\begin{align}\label{gunthers-equation}
\ddot{x}^ag_{ab}+\Gamma_{bcd}\dot{x}^c\dot{x}^d=0.
\end{align}

Two remarkable comments are now in order. First of all, the above equation pertains to extremals of the metric $g_{ab}$ as it can be derived  via an action principle with Lagrangian   $\mathcal{L}=\left( g_{ab}\dot{x}^a\dot{x}^b\right)^{1/2}$, similar to standard scenarios (notice that since  $\nabla^*$ preserves the metric, Eq. \eqref{constant-lagrangian} is still valid; equivalently, this identity follows from the fact that  the Lagrangian is constant along curves satisfying Eq. \eqref{gunthers-equation}). Moreover, Eq. \eqref{gunthers-equation}  represents what is referred to as \qm{geodesic} in the research field devoted to the analysis of the so-called spacetime defects (see Eq. (3.1.14) in Ref. \cite{Gunther}). These objects are related to intriguing geometries whose  metric becomes degenerate on some hypersurface embedded in the spacetime manifold, and have  been recently employed in the context of  bouncing cosmologies \cite{Klinkhamer-Wang2019,Wang2021,Battista2020}, wormholes \cite{Klinkhamer2022}, and string theory \cite{Klinkhamer2021}. We can thus say that our study has  provided a formal and solid characterization of these \qm{geodesic curves}.

A priori, and contrary to Eq. \eqref{autoparallel-eq1}, Eq. \eqref{gunthers-equation} is not well-posed in general. Indeed, the former is written in normal form (i.e., the second-derivative quantity is isolated and its   coefficient is one) and hence the existence theorem (as well as the explosion alternative theorem mentioned in Sec. \ref{Sec:geodesic-completeness}) holds. On the other hand, in Eq. \eqref{gunthers-equation},  it is generally impossible to \qm{invert} the second-derivative contribution and express it in terms of the first-derivative piece. This implies, as first discussed in Refs. \cite{Gunther,Klinkhamer-Wang2019}, that  singularity theorems can  typically be evaded  in spacetimes featuring a degenerate metric. Remarkably,  this does not happen for  the Schwarzschild solution   \eqref{Schwarzschild-metric-matrix-form}, whose special geometric  properties   permit to isolate the second-derivative factors  of Eq. \eqref{gunthers-equation}, which thus turns out to be  well-posed (see Eqs. \eqref{temporal-gunther-all} and \eqref{radial-gunthers} below).

By limiting our attention to the equatorial slice of the Schwarzschild spacetime, we find that the temporal components of Eq. \eqref{gunthers-equation} are 
\begin{subequations}
\label{temporal-gunther-all}
\begin{align}
\ddot{T}&=\frac{2M\dot{r}\dot{T}}{r(2M-r)},
\label{temporal-gunther-1} \\
\ddot{t}&=\frac{2M\dot{r}\dot{t}}{r(2M-r)}, 
\label{temporal-gunther-2} \\
\dot{T}\dot{t}&=0. \label{temporal-gunther-3}
\end{align}
\end{subequations}
The last equation indicates that each contravariant autoparallel curve has at least one temporal coordinate remaining constant.  In addition, the equation for $\phi$ is again given by Eq. \eqref{for-phi}, while for the radial component we obtain
\begin{align}
\ddot{r}=\left(1-\frac{2M}{r}\right)\left\{r\dot{\phi}^2+\frac{M}{r^2}\left[\frac{\dot{r}^2}{(1-2M/r)^2}+\dot{T}^2-\dot{t}^2\right]\right\},
\label{radial-gunthers}
\end{align}
which amounts to be exactly the real part of Eq. \eqref{for-r-Schwarzschild-1}.

Therefore, the angular and radial equations match  those of the Lorentzian (resp. Euclidean) Schwarzschild geometry when $\dot{T}=0$ (resp. $\dot{t}=0$). In other words, up to temporal translations, the contravariant  autoparallels  live entirely in either  the Euclidean or   Lorentzian section of the complex Schwarzschild spacetime. Being allowed to move in the Lorentzian section of the spacetime, these curves  can reach the $r=0$ singularity in a finite amount of proper time. In addition, the trajectories hitting the singularity  always preserve their Lorentzian character during their entire evolution. This suggests the  motto \qm{once Lorentzian, always Lorentzian}, which resembles  the famous statement of Ref. \cite{MTW} \qm{once timelike, always timelike}. The same behaviour is valid for Euclidean curves, which keep their nature unaltered throughout the dynamics.  As we have demonstrated in Sec. \ref{Sec:geodesic-completeness}, in this case the $r=0$ singularity is never attained. We also notice that, in our analysis, both Eqs. \eqref{autoparallel-schwarzschild} and \eqref{gunthers-equation} give  no new orbits that have not already been encountered in Euclidean and Lorentzian Schwarzschild geometries. This ties in with the fact that all the curves analyzed here are real. 

As a final remark,  we deem that the extremals, or equivalently the contravariant autoparallels, discussed in this section can  physically represent the degenerate-case analogue of the geodesics for at least three reasons. Firstly, their equations stem from an  action principle; secondly, they have been widely explored in the 
literature,  showing that they can explain important phenomena such as nonsingular bouncing models and have crucial implications for singularity theorems (for further details, see Ref. \cite{Wang-thesis} and references therein); lastly, they exhibit a richer structure than the autoparallels considered in Sec. \ref{Sec:autoparallels}, since  static curves only represent a particular class of solutions of Eq. \eqref{temporal-gunther-all}.

\section{Conclusions}\label{Sec:conclusions}

Complex metrics are extensively employed in both quantum gravity, where they provide convergent integration contours for the gravitational path integral, and in classical contexts, particularly for their potential to avoid singularities. Inspired by the fact that dealing with degenerate metrics is often required in these settings,   in this paper we have proposed a covariant generalization of the Moore-Penrose procedure, which allows  general relativity to be extended to cases involving nowhere-invertible metrics.  In this approach, the uniqueness of the pseudoinverse metric arises from a set of covariant relations that generalize the standard Moore-Penrose formulas. This new formalism permits the introduction of a covariant derivative operator, and the unambiguous definition of curvature-related tensors (i.e., Riemann, Ricci, and Einstein tensors) and scalars (i.e.,  Kretschmann invariant). Remarkably, the degenerate nature of the metric is reflected in the presence of a torsion tensor,  which thus has a purely geometric origin. 

One of the main advantages of the proposed covariant inversion algorithm is its suitability for both real-valued and complex-valued degenerate tensors of any constant rank  in any spacetime dimension. We have considered some applications by examining both   static black-hole geometries and cosmological models. These time-complexified five-dimensional degenerate metrics satisfy the Einstein equations  and can  thus be interpreted as  generalized instantons  exhibiting no \qm{preferred direction} of time,  which we  propose to  term \qm{arrowless instantons}. In this regard, it is worth noting that an important aspect that needs further exploration concerns the background Riemannian metric $\zeta_{ab}$ that enters the covariant inversion  process. While no natural choice of $\zeta_{ab}$ seems  immediately evident, we have proved  that when the metric $g_{ab}$ commutes with $\zeta_{ab}$ in some coordinate chart,  $\tilde{g}^{ab}$ boils down to the  conventional Moore-Penrose pseudoinverse. Specifically, in the analysis of the aforementioned \qm{arrowless instantons}, $\zeta_{ab}$ is flat and hence  does not contribute to the spacetime curvature. This suggests that taking  $\zeta_{ab}$ as simple as possible can yield  sensible results  that generalize the usual ones. The extent to which physical predictions  depend on the choice of $\zeta_{ab}$ is an important topic that deserves consideration in a separate paper.

By adopting our novel inversion strategy, we have  introduced  a broader concept of geodesics that builds on two  types of curves, namely autoparallels and extremals. Using the complex Schwarzschild metric, which has represented  our reference solution throughout the paper, we have found that extremals, here referred to as \qm{contravariant autoparallels},  can physically represent the degenerate-case analogue of geodesics. These trajectories evolve entirely within either  the Lorentzian or Euclidean section of the complexified spacetime. In the former case, they can reach the singularity at $r=0$, while in the latter they cannot, since the Euclidean domain of the complexified Schwarzschild spacetime is geodesically complete  (an original proof of this result  has been given in Sec. \ref{Sec:geodesic-completeness}).

Our findings provide promising starting points for further research in both classical and quantum gravity. First of all, the aforementioned generalized geodesics could play a key role in the examination of spacetime singularities, as the analysis of their behaviour can reveal whether the complexified spacetime is singular or not. Furthermore, the complex Schwarzschild geometry studied in this paper comprises both the Lorentzian and  Euclidean domains and hence can describe a spacetime with varying topology. As explained in Ref. \cite{dowker-surya}, the fact that the underlying three-dimensional space of a classical spacetime has ``frozen'' topology leads to issues in quantum gravity. This explains why topology changes are considered: following Ref. \cite{dowker-surya}, a topology change is a Lorentzian cobordism between two non-diffeomorphic Riemannian manifolds of the same dimension, called the \emph{initial} and \emph{final} spaces. In our paper, we have a similar situation: the complexified five-dimensional manifolds $\mathcal{M}$ we have examined can be interpreted as cobordisms between Euclidean and Lorentzian domains. To be more precise, the (non-connected) sub-manifold $\{T+\ii t\ne0\}$ of $\mathcal{M}$ is a cobordism between the subspace $\{T=0,~t\ne0\}$ of the Lorentzian section, and the subspace $\{t=0,~T\ne0\}$ of the Euclidean domain. This cobordism carries a degenerate metric, and it could be an interesting matter of future study to look at the topological properties of such a cobordism, in order to find new bridges between Euclidean and Lorentzian sections of classical spacetimes.

In view of our covariant inversion scheme, complex degenerate metrics might be included in the  gravitational path integral, with potentially important implications for a functional-integral approach to quantum gravity. Moreover, following our analysis of the complex FLRW paradigm, significant consequences can unfold  in quantum cosmology. In this framework,  one can  suppose that our current Lorentzian universe has originated from a five-dimensional time-complexified FLRW spacetime whose properties are influenced by the holomorphy of the scale factor, which regulates the presence of torsion (see Sec. \ref{Sec:application-FLRW}). In addition, since existing models of black holes in the literature make use of complexified degenerate metrics and quantum-cosmology-inspired concepts (see e.g. Ref. \cite{Capozziello2024}), far-reaching effects can also arise in black hole physics.

Last but not least, the extended Moore-Penrose procedure can be invoked not only in gravity research but also in any  area where the ordinary Moore-Penrose method is typically called for, ranging from pure mathematics and physics to broader contexts (see Refs. \cite{Arfken1995,Strang2014,Baksalary2021}).

\section*{Acknowledgements}

E. B. acknowledges the support of INFN, {\it iniziative specifiche}  MOONLIGHT2, and  thanks  Dr. R. Marotta for fruitful and stimulating discussions.

\appendix

\section{Existence and uniqueness of the generalized Moore-Penrose pseudoinverse  }\label{Sec:technical-stuff}

This is a rather technical section where we prove that  the generalized pseudoinverse $\tilde{g}$, derived through the covariant Moore-Penrose algorithm introduced in Sec. \ref{Sec:pseudoinverse-metric}, exists and is unique.

We begin by  showing  that the right and left dagger operations defined  in Eqs. \eqref{right-dagger} and \eqref{left-dagger} are intrinsic (i.e., coordinate-free). These can be seen  as the maps 
\begin{subequations}
\label{adjoints}
\begin{align}
&\cdot^\dagger : \Gamma{\rm End}(T^*_{\C}\mathcal{M}) \longrightarrow \Gamma{\rm End}(T^*_{\C}\mathcal{M}), \label{adjoints-1} \\
&{}^\dagger{\cdot} : \Gamma{\rm End}(T_{\C}\mathcal{M}) \longrightarrow \Gamma{\rm End}(T_{\C}\mathcal{M}), \label{adjoints-2}
\end{align}
\end{subequations}
where $\Gamma E$ denotes the space of global sections of a (complex) vector bundle $E$ on $\mathcal{M}$, and ${\rm End}(V) \equiv {\rm Hom}(V,V)$, ${\rm Hom}(V,V')$ being the bundle of homomorphisms between the vector bundles $V$ and $V'$. The maps \eqref{adjoints} are constructed as follows. First of all, the transposition endomorphism  \cite[Chap. 5, \S 6]{husemoller}
\begin{align}\label{transposition}
&{}^t{\cdot} : \Gamma{\rm Hom}(T^*\mathcal{M},T\mathcal{M})\longrightarrow\Gamma{\rm Hom}(T^*\mathcal{M},(T^*\mathcal{M})^*)=\Gamma{\rm Hom}(T^*\mathcal{M},T\mathcal{M}),
\end{align}
tensored with complex conjugation yields a complex-transpose map
\begin{align}\label{conjugate-transpose}
    &{\cdot}^*={}^t{\overline{\cdot}} : \Gamma{\rm Hom}(T^*_\C\mathcal{M},T_\C\mathcal{M})\longrightarrow\Gamma{\rm Hom}(T^*_\C\mathcal{M},T_\C\mathcal{M}).
\end{align}
On the other hand,  the index-raising and index-lowering morphisms attached to the base Riemannian metric $\zeta$ 
\begin{align}\label{sharp-and-flat}
    \begin{array}{ccccc}\sharp & : & T^*\mathcal{M} & \longrightarrow & T\mathcal{M}, \\ & & v_a & \longmapsto & \zeta^{ab}v_b,\end{array}~~\text{and}~~\begin{array}{ccccc}\flat & : & T\mathcal{M} & \longrightarrow & T^*\mathcal{M}, \\ & & v^a & \longmapsto & \zeta_{ab}v^b,\end{array}
\end{align}
induce the maps 
\begin{align}\label{hsharp-and-hflat}
    \begin{array}{ccc}\Gamma{\rm End}(T^*\mathcal{M}) & \stackrel{\sharp}\longrightarrow & \Gamma{\rm Hom}(T^*\mathcal{M},T\mathcal{M}), \\ {X_a}^b & \longmapsto & \zeta^{ac}{X_c}^b,\end{array}~~\text{and}~~\begin{array}{ccc}\Gamma{\rm Hom}(T^*\mathcal{M},T\mathcal{M}) & \stackrel{\flat}\longrightarrow & \Gamma{\rm End}(T^*\mathcal{M}), \\ Y^{ab} & \longmapsto & \zeta_{ac}Y^{cb}.\end{array}
\end{align}
Finally, we let
\begin{align}\label{dagger-up}
    \begin{array}{ccccc}\cdot^\dagger\equiv(\flat)\circ(\cdot^*)\circ(\sharp) & : & \Gamma{\rm End}(T^*_\C\mathcal{M}) & \longrightarrow & \Gamma{\rm End}(T^*_\C\mathcal{M}), \\ & & {X_a}^b & \longmapsto & \zeta_{ac}{\overline{X}_d}^c\zeta^{db},\end{array}
\end{align}
and, dually,
\begin{align}\label{dagger-down}
    \begin{array}{ccccc}{}^\dagger{\cdot} & : & \Gamma{\rm End}(T_\C\mathcal{M}) & \longrightarrow & \Gamma{\rm End}(T_\C\mathcal{M}), \\ & & {Y^a}_b & \longmapsto & \zeta^{ac}{\overline{Y}^d}_c\zeta_{db}.\end{array}
\end{align}
The above formulas thus show that the right and left dagger operations are intrinsic constructions, since they are defined using only intrinsic objects (namely,  morphisms between vector bundles), as claimed before. 

We now prove that, given a Riemannian manifold $(\mathcal{M},\zeta)$ and a symmetric $(2,0)$-tensor $g$ of constant rank on $\mathcal{M}$, there exists a unique $(0,2)$-tensor $\tilde{g}$ satisfying the generalized covariant Moore-Penrose conditions \eqref{Moore-Penrose-metrici} (or equivalently \eqref{Moore-Penrose-metricc}).

The proof  will be done by gluing local pseudoinverses on charts. Thus, we first solve the case where $\mathcal{M}=U\subset\R^n$ is an open submanifold of $\R^n$. Then, the metric $\zeta$ is given in coordinates by $\zeta=\zeta_{ab}\dd x^a\otimes \dd x^b$ ($a,b=1,\dots,n$). Assume first that $\zeta_{ab}=\delta_{ab}$ i.e. $\zeta=(\dd x^1)^2+\cdots+(\dd x^n)^2$ is the standard inner product on $\R^n$. We have to prove that the map $x\mapsto \tilde{g}^{ab}(x)$, where $\tilde{g}^{ab}(x):=\widetilde{g_{ab}(x)}$ is the classical Moore-Penrose inverse of the matrix $g_{ab}(x)$, is smooth. Recall that since the assignment $x\mapsto g_{ab}(x)$ is continuous and since $x\mapsto{\rm rk}(g_{ab}(x))$ is constant (remember that we have assumed $g$ to have constant rank on $\mathcal{M}$), the fundamental result of Penrose \cite{Penrose1955} ensures that the map $x\mapsto \tilde{g}^{ab}(x)$ is continuous. Now, using \cite[Prop 3.15]{Hjorungnes} and again the fact that $g_{ab}$ has constant rank, we find that $x\mapsto \tilde{g}^{ab}(x)$ is differentiable and its partial derivatives are given by 
\begin{align}
    (\partial_u\tilde{g})^{ab}=-\tilde{g}^{ac}(\partial_ug_{cd})\tilde{g}^{db}+\tilde{g}^{ac}{\tilde{g}^*}_{cd}(\partial_u{g^*}^{de})({\delta_e}^b-g_{ec}\tilde{g}^{cb})+({\delta^a}_c-\tilde{g}^{ae}g_{ec})(\partial_u{g^*}^{cd}){\tilde{g}^*}_{de} \tilde{g}^{eb},
\end{align}
where all indices run from $1$ to $n$. 
Applying Penrose's result again, this implies that $\partial_u\tilde{g}$ is continuous, so that $x\mapsto\tilde{g}^{ab}(x)$ is of class $\mathcal{C}^1$. By induction on this argument, we obtain that $x\mapsto\tilde{g}^{ab}(x)$ is smooth on $U$. Moreover, the uniqueness of the Moore-Penrose inverse ensures that the tensor field $\tilde{g}^{ab}$ is unique.

We now treat the case of a general metric $\zeta$ on $U \subset\R^n$. Recall from \cite[Theorem 4.2.5]{Saerkkae} that for all $x\in U$, there is a unique upper-triangular matrix $B(x)$ with positive diagonal elements yielding the Cholesky decomposition   $\zeta(x)={}^t{B(x)}B(x)$ (see also footnote \ref{footnote-1}). Recall from \cite[Lemma 12.1.6]{Schatzman} that the map $x\mapsto B(x)$ is continuous and using now \cite[Theorem A.1]{Golub-Loan}, the map $x\mapsto B(x)$ is of class $\mathcal{C}^1$, with partial derivatives 
\begin{align}\label{eq:smooth_cholesky}
    \partial_uB=\Psi({}^t{B^{-1}}(\partial_u\zeta)B^{-1})B,
\end{align}
where the function $\Psi$ returns the strict upper triangular part plus half the diagonal part of a matrix $A$, i.e. 
\begin{align}
    \Psi(A)_{ab}=\left\{\begin{array}{cc}
A_{ab} & \text{if $a<b$}, \\ \tfrac12 A_{ab} & \text{if $a=b$}, \\ 0 & \text{if $a>b$}.\end{array}\right.
\end{align}
Inductively again, we see that $x\mapsto B(x)$ is in fact smooth. Observe now that $\tilde{g}$ is a pseudoinverse for $g$ with respect to $\zeta$ if and only if $\tilde{g}_B:=B\tilde{g}{}^tB$ is a pseudoinverse for $g_B:={}^tB^{-1}gB^{-1}$ with respect to the standard metric on $\R^n$. %Indeed, \[g\tilde{g}g=g~\Leftrightarrow~{}^tB^{-1}gB^{-1}B\tilde{g}{}^tB{}^tB^{-1}gB^{-1}={}^tB^{-1}gB^{-1}~\Leftrightarrow~g_B\tilde{g}_Bg_B=g_B\] and similarly, $\tilde{g}g\tilde{g}=\tilde{g}~\Leftrightarrow~\tilde{g}_Bg_B\tilde{g}_B=\tilde{g}_B$. Next, because \[(g\tilde{g})^\dagger=\eta(g\tilde{g})^*\eta^{-1}={}^tBB({}^tB(g_B\tilde{g}_B){}^tB^{-1})^*B^{-1}{}^tB^{-1}={}^tBBB^{-1}(g_B\tilde{g}_B)^*BB^{-1}{}^tB^{-1}={}^tB(g_B\tilde{g}_B)^*{}^tB^{-1}\] we get \[(g\tilde{g})^\dagger=g\tilde{g}~\Leftrightarrow~{}^tB(g_B\tilde{g}_B)^*{}^tB^{-1}=g\tilde{g}={}^tB(g_B\tilde{g}_B){}^tB^{-1}~\Leftrightarrow~(g_B\tilde{g}_B)^*=g_B\tilde{g}_B\] and similarly, $(\tilde{g}g)_\dagger=\tilde{g}g~\Leftrightarrow~(\tilde{g}_Bg_B)^*=\tilde{g}_Bg_B$.
Therefore, the pseudoinverse $\tilde{g}=B^{-1}\tilde{g}_B{}^tB^{-1}$ exists, is unique, and is smooth, because $x\mapsto \tilde{g}_B(x)$ and $x\mapsto B(x)^{-1}$ are smooth, as demonstrated above.

Finally, if $(\mathcal{M},\zeta)$ is a general (connected) Riemannian manifold of dimension $n$, we consider an atlas $ \left\{U_\alpha, \varphi_\alpha \right\} $  of $\mathcal{M}$. On the open subset $\varphi_\alpha(U_\alpha)\subset\R^n$, there exists a unique Moore-Penrose inverse $\hat{g}_\alpha$ for $(\varphi_\alpha^{-1})^*(g_{|U_\alpha})$ with respect to $(\varphi_\alpha^{-1})^*(\zeta_{|U_\alpha})$. By covariance, the local section $\tilde{g}_\alpha:=\varphi_\alpha^*(\hat{g}_\alpha)$ is a Moore-Penrose inverse for $g_{|U_\alpha}$ relatively to $\zeta_{|U_\alpha}$. Because of the covariance and uniqueness of the pseudoinverse, if $\alpha,\beta$ are such that $U_\alpha\cap U_\beta\ne\emptyset$, then $(\tilde{g}_\alpha)_{|U_\alpha\cap U_\beta}=(\tilde{g}_\beta)_{|U_\alpha\cap U_\beta}$. Therefore, we can glue these local sections of $T^*_\C\mathcal{M}\otimes T^*_\C\mathcal{M}$ together\footnote{In mathematical terms, we say that the local sections form a {\it sheaf}.}, to form a unique global section $\tilde{g}\in\Gamma \left(T^*_\C\mathcal{M}\otimes T^*_\C\mathcal{M}\right)$ such that $\tilde{g}_{|U_\alpha}=\tilde{g}_\alpha$, with $\tilde{g} $ a symmetric tensor. This is the desired covariant Moore-Penrose inverse of $g$ relatively to $\zeta$, finishing the proof.

\section{Existence and uniqueness of the $g$-invariant connection }\label{Sec:unique-connection}

We prove here that the connection \eqref{def-nabla} is the only affine connection on $\mathcal{M}$ satisfying the following properties 
\begin{subequations}
\begin{align}
    \nabla g=0, \label{add-conds-3} \\
    \tilde{g}(\nabla^*-g\nabla)\tilde{g}=0, \label{add-conds-1} \\
    \tilde{g}g(\nabla-\hat\nabla)=\nabla-\hat\nabla, \label{add-conds-2}
\end{align}
\label{add-conds}
\end{subequations}
where $\nabla^*$ is the contravariant connection introduced in Sec. \ref{Sec:Extremal-curves} and $\hat\nabla$ is the Levi-Civita connection associated to the Riemannian metric $\zeta_{ab}$. The condition \eqref{add-conds-3}  expresses the $g$-invariance of the connection $\nabla$, while the conditions \eqref{add-conds-1} and \eqref{add-conds-2} are to be understood as equalities between tensors, which means that  these two equations are intrinsic. In components, the set of relations \eqref{add-conds} assume the form  
\begin{subequations}
\begin{align}
    g_{ad}{\Gamma^d}_{cb}+g_{bd}{\Gamma^d}_{ca}=\partial_cg_{ab}, \label{conds-comp-1} \\
    \tilde{g}^{ab}(\Gamma_{bcd}-g_{bu}{\Gamma^u}_{cd})\tilde{g}^{de}=0, \label{conds-comp-2} \\
    \tilde{g}^{ab}g_{bc}({\Gamma^c}_{de}-{\hat\Gamma^c}_{de})={\Gamma^a}_{de}-{\hat\Gamma^a}_{de}. \label{conds-comp-3}
\end{align}
\label{conds-comp}
\end{subequations}
Now, we have to prove that the connection coefficients ${\Gamma^{a}}_{bc}$, which we have derived in Eq. \eqref{second-christoffel}, are the only ones satisfying  the above identities.

The last equation can be rewritten as
\begin{align}
    \tilde{g}^{ab}g_{bc}{\Gamma^c}_{de}-{\Gamma^a}_{de}=\tilde{g}^{ab}g_{bc}{\hat\Gamma^c}_{de}-{\hat\Gamma^a}_{de},
\end{align}
and hence raising the last index $e$ and using Eq. \eqref{conds-comp-2}, we arrive at
\begin{align}
    \tilde{g}^{ab}\Gamma_{bde}\tilde{g}^{ef}-{\Gamma^a}_{de}\tilde{g}^{ef}=(\tilde{g}^{ab}g_{bc}{\hat\Gamma^c}_{de}-{\hat\Gamma^a}_{de})\tilde{g}^{ef}.
\end{align}
Upon invoking this formula jointly with  Eq. \eqref{conds-comp-1}, we get
\begin{align}
    \tilde{g}^{ad}\partial_bg_{dc}&=\tilde{g}^{ad}g_{de}{\Gamma^e}_{bc}+g_{ce}\tilde{g}^{ad}{\Gamma^e}_{bd} \nonumber\\
    &=\tilde{g}^{ad}g_{de}{\hat\Gamma^e}_{bc}-{\hat\Gamma^a}_{bc}+{\Gamma^a}_{bc}+g_{ce}\left[\tilde{g}^{ef}\Gamma_{fbd}-\tilde{g}^{ef}g_{fh}{\hat\Gamma^h}_{bd}+{\hat\Gamma^e}_{bd}\right]\tilde{g}^{da} \nonumber\\
    &={\Gamma^a}_{bc}+g_{ce}\tilde{g}^{ef}\Gamma_{fbd}\tilde{g}^{da}-{\hat\Gamma^a}_{bc}+g_{de}{\hat\Gamma^e}_{bc}\tilde{g}^{da},
\end{align}
and thus
\begin{align}
    {\Gamma^a}_{bc}&=\tilde{g}^{ad}\partial_bg_{dc}-g_{ce}\tilde{g}^{ef}\Gamma_{fbd}\tilde{g}^{da}+{\hat\Gamma^a}_{bc}-g_{de}{\hat\Gamma^a}_{bc}\tilde{g}^{da} \nonumber\\
    &=\tilde{g}^{ad}(\Gamma_{dbc}+\Lambda_{cbd})+{\hat\Gamma^a}_{bc}-\tilde{g}^{ad}g_{de}{\hat\Gamma^e}_{bc},
\end{align}
which is exactly Eq. \eqref{second-christoffel}. Conversely, if we  start with the above expression for $ {\Gamma^a}_{bc}$, then it is simple to prove that it satisfies the required properties \eqref{conds-comp}.

It is important, though trivial, to notice that if $g_{ab}$ is non-degenerate, the only connection satisfying the conditions \eqref{add-conds} is the Levi-Civita connection.

\section{ Riemann tensor properties }\label{Sec:properties-proofs}

In this section we demonstrate the identities satisfied by the Riemann tensor, which has been introduced in Eq. \eqref{contra-riemann-coefs}. 

Let us start with the first Bianchi identity \eqref{bianchi-I}. Bearing in mind the definitions   $R(X,Y):=[\nabla_X,\nabla_Y]-\nabla_{[X,Y]}$ and $ T(X,Y):=\nabla_XY-\nabla_YX-[X,Y]$, we can calculate 
\begin{align}
3R(\partial_{[b},\partial_{c})\partial_{d]}=&R(\partial_b,\partial_c)\partial_d+R(\partial_c,\partial_d)\partial_b+R(\partial_d,\partial_b)\partial_c \nonumber\\
    =&\nabla_b\nabla_c\partial_d-\nabla_c\nabla_b\partial_d+\nabla_c\nabla_d\partial_b-\nabla_d\nabla_c\partial_b+\nabla_d\nabla_b\partial_c-\nabla_b\nabla_d\partial_c \nonumber\\
    =&\nabla_b(\nabla_c\partial_d-\nabla_d\partial_c)+\nabla_c(\nabla_d\partial_b-\nabla_b\partial_d)+\nabla_d(\nabla_b\partial_c-\nabla_c\partial_b) \nonumber\\
=&\nabla_b({T_{cd}}^k\partial_k)+\nabla_c({T_{db}}^k\partial_k)+\nabla_d({T_{bc}}^k\partial_k)=3\nabla_{[b}({T_{cd]}}^k\partial_k) \nonumber\\
    =&\left[\partial_b({T_{cd}}^k)+\partial_c({T_{db}}^k)+\partial_d({T_{bc}}^ k)\right]{\partial_k} \nonumber\\
    &+{T_{cd}}^k\nabla_b\partial_k+{T_{db}}^k\nabla_c\partial_k+{T_{bc}}^k\nabla_d\partial_k \nonumber\\
    %=&3\left(\partial_{[b}{T_{cd]}}^k\widecheck{\partial_k}+{T_{[bc}}^k\nabla_{d]}\partial_k\right)=3\left(\partial_{[b}{T_{cd]}}^k+{T_{[bc}}^k\nabla_{d]}\right)\partial_k
    %=&\left[\partial_b({T_{cd}}^v)+{T_{cd}}^k\tilde{g}^{uv}(\Gamma_{ubk}+\Lambda_{kbu})\right]\partial_v \\
    %&+\left[\partial_c({T_{db}}^v)+{T_{db}}^k\tilde{g}^{uv}(\Gamma_{uck}+\Lambda_{kcu})\right]\partial_v \\
    %&+\left[\partial_d({T_{bc}}^v)+{T_{bc}}^k\tilde{g}^{uv}(\Gamma_{udk}+\Lambda_{kdu})\right]\partial_v
    =&\left[\partial_b({T_{cd}}^v)+{T_{cd}}^k(\tilde{g}^{uv}(\Gamma_{ubk}+\Lambda_{kbu})+{\hat\Gamma^v}_{bk}-\tilde{g}^{vu}g_{uw}{\hat\Gamma^w}_{bk})\right]\partial_v \nonumber\\
    &+\left[\partial_c({T_{db}}^v)+{T_{db}}^k(\tilde{g}^{uv}(\Gamma_{uck}+\Lambda_{kcu})+{\hat\Gamma^v}_{ck}-\tilde{g}^{vu}g_{uw}{\hat\Gamma^w}_{ck})\right]\partial_v \nonumber\\
    &+\left[\partial_d({T_{bc}}^v)+{T_{bc}}^k(\tilde{g}^{uv}(\Gamma_{udk}+\Lambda_{kdu})+{\hat\Gamma^v}_{dk}-\tilde{g}^{vu}g_{uw}{\hat\Gamma^w}_{dk})\right]\partial_v.
    \label{relation-app-C-1}
\end{align}
Now, by definition of  covariant derivative of tensors, we have
\begin{align}%\label{def-nabla-T}
    \partial_b({T_{cd}}^v)-{(\nabla_bT)_{cd}}^v=&\left[\tilde{g}^{ku}(\Gamma_{ubc}+\Lambda_{cbu})+{\hat\Gamma^k}_{bc}-\tilde{g}^{ku}g_{uw}{\hat\Gamma^w}_{bc}\right]{T_{kd}}^v \\
    &+\left[\tilde{g}^{ku}(\Gamma_{ubd}+\Lambda_{dbu})+{\hat\Gamma^k}_{bd}-\tilde{g}^{ku}g_{uw}{\hat\Gamma^w}_{bd}\right]{T_{ck}}^v \\
    &-\left[\tilde{g}^{vu}(\Gamma_{ubk}+\Lambda_{kbu})+{\hat\Gamma^v}_{bk}-\tilde{g}^{vu}g_{uw}{\hat\Gamma^w}_{bk}\right]{T_{cd}}^k,
    %\partial_b({T_{cd}}^k)={(\nabla_bT)_{cd}}^k+\tilde{g}^{uv}(\Gamma_{ubc}+\Lambda_{cbu}){T_{vd}}^k+\tilde{g}^{uv}(\Gamma_{ubd}+\Lambda_{dbu}){T_{cv}}^k-\tilde{g}^{kv}(\Gamma_{vbu}+\Lambda_{ubv}){T_{cd}}^u
\end{align}
and thus we get%because of the relation ${T_{cd}}^kg_{kl}\tilde{g}^{lw}={T_{cd}}^w$, we get
\begin{align}
    \hspace{-5mm}{T_{cd}}^k\left[\tilde{g}^{uv}(\Gamma_{ubk}+\Lambda_{kbu})+{\hat\Gamma^v}_{bk}-\tilde{g}^{vu}g_{uw}{\hat\Gamma^w}_{bk}\right]=&\left[\tilde{g}^{ku}(\Gamma_{ubc}+\Lambda_{cbu})+{\hat\Gamma^k}_{bc}-\tilde{g}^{ku}g_{uw}{\hat\Gamma^w}_{bc}\right]{T_{kd}}^v \nonumber\\
    &+\left[\tilde{g}^{ku}(\Gamma_{ubd}+\Lambda_{dbu})+{\hat\Gamma^k}_{bd}-\tilde{g}^{ku}g_{uw}{\hat\Gamma^w}_{bd}\right]{T_{ck}}^v \nonumber\\
    &+{(\nabla_bT)_{cd}}^v-\partial_b({T_{cd}}^v).
    %\partial_b({T_{cd}}^w)+{T_{cd}}^k\tilde{g}^{uw}(\Gamma_{ubk}+\Lambda_{kbu})=&{(\nabla_bT)_{cd}}^w+\cancel{{T_{cd}}^k\tilde{g}^{uw}(\Gamma_{ubk}+\Lambda_{kbu})} \\
    %&+\tilde{g}^{uv}(\Gamma_{ubc}+\Lambda_{cbu}){T_{vd}}^w+\tilde{g}^{uv}(\Gamma_{ubd}+\Lambda_{dbu}){T_{cv}}^w \\
    %&\cancel{-\tilde{g}^{kv}(\Gamma_{vbu}+\Lambda_{ubv}){T_{cd}}^ug_{kl}\tilde{g}^{lw}}.
\end{align}
Plugging this formula into Eq. \eqref{relation-app-C-1}, and  exploiting  the symmetries $\Gamma_{a[bc]}=0$ and ${\hat\Gamma^a}_{[bc]}=0$, we obtain
\begin{align}
    3R(\partial_{[b},\partial_c)\partial_{d]}=&\left[{(\nabla_bT)_{cd}}^w+{(\nabla_cT)_{db}}^w+{(\nabla_dT)_{bc}}^w+(\Lambda_{dcu}-\Lambda_{cdu})\tilde{g}^{uv}{T_{vb}}^w\right. \nonumber\\
%    3R(\partial_{[b},\partial_c)\partial_{d]}=&\left[{(\nabla_bT)_{cd}}^w+\tilde{g}^{uv}(\cancel{\Gamma_{ubc}}+\Lambda_{cbu}){T_{vd}}^w+\tilde{g}^{uv}(\cancel{\Gamma_{ubd}}+\Lambda_{dbu}){T_{cv}}^w\right. \\
%    &+{(\nabla_cT)_{db}}^w+\tilde{g}^{uv}(\cancel{\Gamma_{ucd}}+\Lambda_{dcu}){T_{vb}}^w+\tilde{g}^{uv}(\cancel{\Gamma_{ucb}}+\Lambda_{bcu}){T_{dv}}^w \\
%    &\left.+{(\nabla_dT)_{bc}}^w+\tilde{g}^{uv}(\cancel{\Gamma_{udb}}+\Lambda_{bdu}){T_{vc}}^w+\tilde{g}^{uv}(\cancel{\Gamma_{udc}}+\Lambda_{cdu}){T_{bv}}^w\right]\partial_w \\
    %=&\left[{(\nabla_bT)_{cd}}^w+{(\nabla_cT)_{db}}^w+{(\nabla_dT)_{bc}}^w\right. \\
    &+\left.(\Lambda_{bdu}-\Lambda_{dbu})\tilde{g}^{uv}{T_{vc}}^w+(\Lambda_{cbu}-\Lambda_{bcu})\tilde{g}^{uv}{T_{vd}}^w\right]\partial_w \nonumber\\
    =&\left(3\nabla_{[b}{T_{cd]}}^w+{T_{cd}}^v{T_{vb}}^w+{T_{db}}^v{T_{vc}}^w+{T_{bc}}^v{T_{vd}}^w\right)\partial_w \nonumber\\
    =&3\left(\nabla_{[b}{T_{cd]}}^w+{T_{v[b}}^w{T_{cd]}}^v\right)\partial_w,
\end{align}
which represents   Eq. \eqref{bianchi-I} written in covariant form, i.e., 
\begin{align}\label{bianchi-I-low}
    R_{a[bcd]}=g_{ae}\left(\nabla_{[b}{T_{cd]}}^e-{T_{[bc}}^f{T_{d]f}}^e\right).%+{T_{v[b}}^u{T_{cd]}}^v\right).
\end{align}

We can now prove that  property $R_{abcd}=R_{cdab}$ no longer holds due to the presence of a nonvanishing torsion tensor. By using Eq. \eqref{bianchi-I-low} four times and taking into account that $R_{(ab)cd}=R_{ab(cd)}=0$, we can write
\begin{align}
    2R_{cdab}=&R_{cdab}-R_{cabd}-R_{cbda}+3R_{c[dab]}=R_{dcba}+R_{acbd}+R_{bcda}+3R_{c[dab]} \nonumber\\
    =&R_{dcba}-R_{abdc}-R_{adcb}-R_{bdac}-R_{bacd}+3(R_{c[dab]}+R_{a[bdc]}+R_{b[cda]}) \nonumber\\
    =&2R_{abcd}+3(R_{c[dab]}+R_{a[bdc]}+R_{b[cda]}+R_{d[acb]}) \nonumber\\
    =&2R_{abcd}-3(R_{a[bcd]}-R_{b[cda]}-R_{c[dab]}+R_{d[abc]}).
\end{align}
Replacing each cyclic term using identity \eqref{bianchi-I-low} we arrive at the expression
\begin{align}
 \dfrac{2}{3}   \left(R_{abcd}-R_{cdab}\right)=& g_{au}(\nabla_{[b}{T_{cd]}}^u+{T_{v[b}}^u{T_{cd]}}^v)-g_{bu}(\nabla_{[c}{T_{da]}}^u+{T_{v[c}}^u{T_{da]}}^v) \nonumber\\
    &-g_{cu}(\nabla_{[d}{T_{ab]}}^u+{T_{v[d}}^u{T_{ab]}}^v)+g_{du}(\nabla_{[a}{T_{bc]}}^u+{T_{v[a}}^u{T_{bc]}}^v),
    \label{property-riemann-appendix}
\end{align}
showing that $R_{abcd}$ and $R_{cdab}$ differ. 

The second Bianchi identity \eqref{bianchi-II} is easier to derive. Indeed, exploiting the definition of Riemann tensor in the following calculation (the semicolon denotes covariant differentation):
\begin{align}
    3{R^a}_{b[cd;e]}\partial_a=&3(\nabla_{[e}R)(\partial_c,\partial_{d]})\partial_b \nonumber\\
    =&\nabla_e(R(\partial_c,\partial_d)\partial_b)-R(\nabla_e\partial_c,\partial_d)\partial_b-R(\partial_c,\nabla_e\partial_d)\partial_b-R(\partial_c,\partial_d)\nabla_e\partial_b \nonumber\\
    &+\nabla_c(R(\partial_d,\partial_e)\partial_b)-R(\nabla_c\partial_d,\partial_e)\partial_b-R(\partial_d,\nabla_c\partial_e)\partial_b-R(\partial_d,\partial_e)\nabla_c\partial_b \nonumber\\
    &+\nabla_d(R(\partial_e,\partial_c)\partial_b)-R(\nabla_d\partial_e,\partial_c)\partial_b-R(\partial_e,\nabla_d\partial_c)\partial_b-R(\partial_e,\partial_c)\nabla_d\partial_b \nonumber\\
    =&\left(R(\nabla_e\partial_d-\nabla_d\partial_e,\partial_c)+R(\nabla_c\partial_e-\nabla_e\partial_c,\partial_d)+R(\nabla_d\partial_c-\nabla_c\partial_d,\partial_e)\right)\partial_b \nonumber\\
=&\left(R(T(\partial_e,\partial_d),\partial_c)+R(T(\partial_c,\partial_e),\partial_d)+R(T(\partial_d,\partial_c),\partial_e)\right)\partial_b \nonumber\\
    =&\left({T_{ed}}^uR(\partial_u,\partial_c)+{T_{ce}}^uR(\partial_u,\partial_d)+{T_{dc}}^uR(\partial_u,\partial_e)\right)\partial_b \nonumber\\
    =&\left({T_{ed}}^u{R^a}_{buc}+{T_{ce}}^u{R^a}_{bud}+{T_{dc}}^u{R^a}_{bue}\right)\partial_a \nonumber\\
    =&-3{R^a}_{bu[e}{T_{cd]}}^u\partial_a,
\end{align}
easily yields Eq. \eqref{bianchi-II} in its covariant form, which  reads as 
\begin{align}\label{bianchi-II-low}
    R_{ab[cd;e]}+R_{abu[c}{T_{de]}}^u=0.  %\nabla_{[e}R_{|ab|cd]}+R_{abu[e}{T_{cd]}}^u=0,
\end{align}

\bibliography{references}

\end{document}